\pdfoutput=1
\documentclass[bibyear]{aa}  

\usepackage{natbib}
\bibpunct{(}{)}{;}{a}{}{,} 
\usepackage{graphicx}
\usepackage{longtable}
\usepackage{tabularx}
\usepackage{supertabular}
\usepackage{xspace}
\usepackage{amsmath}
\usepackage{threeparttable}
\usepackage{txfonts}
\usepackage{hyperref}
    \hypersetup{colorlinks=true,allcolors=[rgb]{0,0,1}}
    
\usepackage{amsmath}
\usepackage{subcaption}
\usepackage{caption}

\newcommand{\lya}{\mbox{Lyman-$\alpha$}\xspace} 

\newcommand{\ttpaper}{\mbox{\cite{Thai2023LF}}\xspace}
\newcommand{\igpaper}{\mbox{\cite{goovaerts2023}}\xspace}
\newcommand{\igpaperalt}{\mbox{\citealt{goovaerts2023}}\xspace}
\newcommand{\mm}{\mbox{$\,\mu\mathrm{m}$}\xspace}

\newcommand{\ewlya}{\mbox{$\mathrm{EW_{Ly\alpha}}$}\xspace}

\newcommand{\cg}{\mbox{{\tt CIGALE}}\xspace}

\defcitealias{GdlV2019LAELF}{dLV19}
\defcitealias{GdlV2020LAEfrac}{dLV20}
\defcitealias{richard2021atlas}{R21}
\defcitealias{AC2022LLAMAS}{C22}
\defcitealias{S14MS}{S14}
\defcitealias{P23MS}{P23}
\defcitealias{S17MS}{S17}

\DeclareUnicodeCharacter{2212}{-}
\begin{document} 

   \title{Galaxy main sequence and properties of low-mass Lyman-$\alpha$ Emitters towards reionisation viewed by VLT/MUSE and JWST/NIRCam}


   \author{I. Goovaerts\inst{1}\and R. Pello\inst{1} \and D. Burgarella\inst{1} \and T.T.Thai\inst{1,2,3}\and J. Richard\inst{4}\and A. Claeyssens\inst{5} \and P. Tuan-Anh\inst{2,3} \and R. C. Arango-Toro\inst{1} \and L. Boogaard\inst{6} \and T. Contini\inst{7} \and Y. Guo\inst{4} \and I. Langan\inst{4,8} \and N. Laporte\inst{1} \and M. Maseda\inst{9}}
   \titlerunning{Properties of faint SFGs}

   \institute{Aix Marseille Université, CNRS, CNES, LAM (Laboratoire d’Astrophysique de Marseille), UMR 7326, 13388 Marseille, France\\
              \email{ilias.goovaerts@lam.fr}
         \and     
         Department of Astrophysics, Vietnam National Space Center, VAST, 18 Hoang Quoc Viet, Hanoi, Vietnam
         \and
         Department of Astrophysics, Vietnam National Space Center, Vietnam Academy of Science and Technology, 18 Hoang Quoc Viet, Hanoi, Vietnam
         \and
         Univ Lyon, Univ Lyon1, Ens de Lyon, CNRS, Centre de Recherche Astrophysique de Lyon UMR5574, F-69230, Saint-Genis-Laval, France
        \and
        Department of Astronomy, Oskar Klein Centre, Stockholm University, AlbaNova University Centre, SE-106 91 Stockholm, Sweden
        \and 
        Max Planck Institute for Astronomy, Königstuhl 17, 69117 Heidelberg, Germany
        \and 
        Institut de Recherche en Astrophysique et Planétologie (IRAP), Université de Toulouse, CNRS, UPS, CNES, 31400 Toulouse, France
        \and 
        European Southern Observatory, Karl-Schwarzschild-Str. 2, D-85748, Garching, Germany
        \and
        Department of Astronomy, University of Wisconsin-Madison, 475 N. Charter St., Madison, WI 53706, USA
             }

   \date{Received 18 September 2023; accepted XXX}

 
  \abstract
   {Faint, star-forming galaxies likely play a dominant role in cosmic reionisation. Strides have been made in recent years to characterise these populations at high redshifts ($z>3$). Now for the first time, with JWST photometry beyond 1\mm in the rest frame, we can derive accurate stellar masses and position these galaxies on the galaxy main sequence.}
   {We seek to assess the place of 96 individual \lya emitters (LAEs) selected behind the A2744 lensing cluster with MUSE IFU spectroscopy on the galaxy main sequence. We also compare derived stellar masses to \lya luminosities and equivalent widths to better quantify the relationship between the \lya emission and the host galaxy.}
   {These 96 LAEs lie in the redshift range $2.9<z<6.7$, and their range of masses extends down to $10^6\,\mathrm{M_{\odot}}$ (over half with $\mathrm{M_{\star}}<10^8\,\mathrm{M_{\odot}}$). We use the JWST/NIRCam and HST photometric catalogs from the UNCOVER project, giving us excellent wavelength coverage from $450\,\mathrm{nm}$ to $4.5\,$\mm. We perform SED fitting using \cg, fixing the redshift of the LAEs to the secure, spectroscopic value. This combination of photometric coverage with spectroscopic redshifts allows us to robustly derive stellar masses for these galaxies.}
   {We find a main sequence relation for these low mass LAEs of the form: $\mathrm{log\,SFR}=(0.88\pm0.07 - 0.030\pm0.027\times t)\,\mathrm{log\,M_{\star}} - ( 6.31\pm0.41 - 0.08\pm0.37\times t)$. This is in approximate agreement with best-fits of previous collated studies, however, with a steeper slope and a higher normalisation. This indicates that low-mass LAEs towards the epoch of reionisation lie above typical literature main sequence relations derived at lower redshift and higher masses. Additionally, comparing our results to UV-selected samples, we see that while low-mass LAEs lie above these typical main sequence relations, they are likely not singular in this respect at these masses and redshifts. While low-mass galaxies have been shown to play a significant role in cosmic reionisation, our results point to no special position for LAEs in this regard.}
  {}
   
   \keywords{galaxies: high redshift --
                galaxies: evolution --
                gravitational lensing: strong
               }

   \maketitle
%

\section{Introduction}
\label{sect:intro}
\indent The epoch of reionisation is a critical period in the history of the Universe, during which the neutral hydrogen in the intergalactic medium was reionised by the first generations of galaxies, ending around $z\sim5.5$, 1 billion years after the Big Bang \citep{mcgreer2015reionisation,Bouwens2015planckreionisation,Planck2018reionisation,Bosman2022z5.3reionisation}. The sources contributing to reionisation are the subject of ongoing debate in the literature with consensus forming around the importance of faint star-forming galaxies (SFGs) \citep{Finkelstein_2015,Livermore_2017,bouwens2021UVLF,JiangAGN2022}, however, the importance of massive galaxies \cite{Mathee2022brightlya} and active galactic nuclei is still discussed \citep{Grazian2018AGN,Kokorev2023AGN,Fujimoto2023AGNreionisation}.\\
\indent Among SFGs, the particular importance of \lya emitters (galaxies that display \lya emission at $1215.67\,\AA$ rest frame) (henceforth LAEs) has been hypothesised to range from significant \citep{GdlV2019LAELF,Thai2023LF}, to dominant \citep{drake2017MUSELAELF}. LAEs tend to be highly star-forming, young, low-metallicity, dust-poor galaxies (\citealt{stark2010keckLAEfrac,stark2011LAEfrac,deBarros2017LAEfraction}; \igpaperalt). They have been used extensively as probes of the intergalactic medium (IGM) neutral fraction and therefore the progression of reionisation (see, in addition to the studies above: \citealt{stark2010keckLAEfrac,pentericci2011laefrac/z=7LBG,arrabal2018LAEfrac,pentericci2018LAEfrac,caruana2018,GdlV2020LAEfrac,kusakabe2020,bolan2022neutralfraction}). \\
\indent Further constraining the properties of these particular galaxies is crucial, considering their likely importance to the process of reionisation \citep{bouwens2015UVLF,bouwens2022LF_II:LFs}. Now, with \textit{JWST}/NIRCam data pushing wavelength coverage further into the infrared, to $4.5$\mm, which extends into the rest frame optical emission for galaxies between redshifts of 3 and 7, we can obtain more robust estimates for the stellar mass of these high-redshift SFGs.\\
\indent In order to access the intrinsically faint regime of galaxies at redshifts between 3 and 7, gravitational lensing by clusters of galaxies is a useful, and often necessary, tool. Using the magnification from lensing, LAEs have been detected at these redshifts with \lya luminosities as low as $\mathrm{L_{Ly\alpha}\sim10^{39}\,erg\,s^{-1}}$ \citep{richard2021atlas,AC2022LLAMAS,Thai2023LF} and UV-selected sources have been detected down to $\mathrm{M_{UV}}\sim-14$ \citep{GdlV2020LAEfrac,bouwens2022LF_I:sample_selection,goovaerts2023}. Using gravitational lensing and the wavelength coverage of NIRCam now allows us to assess the place of intrinsically faint, very low-mass LAEs on the galaxy main sequence (henceforth MS) in a robust manner. \\
\indent The MS (at low mass) is a linear and typically tight ($\sim0.3$ dex) relation between the stellar mass and star formation rate of galaxies.
Between $10^8<\mathrm{M_{\star}/M_{\odot}}<10^{12}$ and $0<z<6$ the MS is well understood, even taking into account calibrations between different studies related to chosen initial mass function (IMF), chosen cosmology, selection biases and SFR indicator (see primarily the reviews of \citealt{S14MS} (\citetalias{S14MS}) and \citealt{P23MS} (\citetalias{P23MS}), as well as \citealt{brinchmann2004MSlowz,daddi2007MS,elbaz2007MS,santini2009MS,elbaz2011goodsMS,sobral2014stellar,steinhardt2014MS}). In \citetalias{S14MS} a time (or, equivalently, redshift)-dependent form of the MS is suggested. The authors find that the normalisation and slope of the MS decrease with increasing time (decreasing redshift). This result is also found in the review of \citetalias{P23MS}.\\
\indent The position of galaxies on the MS gives an indication of what kind of star-formation a given galaxy is undergoing. A galaxy residing above the typical MS is a starburst galaxy and a galaxy residing below the typical MS is a quenched galaxy with little to no ongoing star formation.\\
\indent Previous studies have made it possible to study the low-mass end of the galaxy MS, using lensing, such as \citet{santini2009MS,S17MS} and at low redshift, such as \citet{boogaard2018low_mass_MS}. The high-redshift studies however, relied on \textit{Spitzer/IRAC} for their coverage past $1600\,\AA$. More recently, \cite{Looser2023MS}, in a sample of 200 galaxies observed with \textit{JWST}/NIRSpec, included estimations of the MS relation down to a very low mass range ($\sim10^{6}\,\mathrm{M_{\odot}}$), in redshift ranges $2<z<5$ and $5<z<11$, finding the galaxies to be located well above the best estimations of the MS at these redshifts and masses. This study, as well as \cite{Atek2023arxiv_lowmassgalaxies} and \cite{Maseda2023NIRSpec_Halpha}, clearly demonstrate \textit{JWST}/NIRCam and \textit{JWST}/NIRSpec's ability to probe these extremely low masses at these redshifts. \cite{Atek2023arxiv_lowmassgalaxies}, in particular, study lensed, reionisation-era galaxies, looking to assess the contribution of intrinsically faint, low-mass galaxies to reionisation. The authors manage to derive stellar masses for 8 such galaxies, down to $\mathrm{log(M_{\star}/M_{\odot}})\sim5.9$, and find through the study of their ionising properties that these galaxies likely drive the bulk of reionisation. \\
\indent In this work, we present a first derivation of this MS relation for intrinsically-faint, low-mass, lensed LAEs towards the epoch of reionisation ($2.9<z<6.7$). We perform our LAE selection blindly using an Integral Field Unit (IFU): MUSE/VLT, meaning we do not suffer from a biased selection where only the UV-brightest sources are considered. Lensing also helps us to remedy this bias, as well as allowing us to access the intrinsically faint and low-mass population dominating cosmic reionisation.\\
\indent Section~\ref{sect:data} presents the spectroscopic and photometric data used. Section~\ref{sect:cigale} covers the details of the SED fitting using secure spectroscopic redshifts from the \lya line. Section~\ref{sect:results} presents the results of placing these faint, low-mass LAEs on the galaxy MS, compares to previous best estimates of the MS relation, and offers conclusions pertinent to these galaxies which likely played a significant role in cosmic reionisation. \\
\indent Throughout this paper, the Hubble constant used is $H_0=70\,\mathrm{km\,s^{-1}Mpc^{-1}}$, and the cosmology is $\Omega_{\Lambda}=0.7$ and  $\Omega_{\mathrm{m}}=0.3$. All values of luminosity and UV magnitude are given corrected for magnification and the IMF adopted is that of \cite{salpeter1955IMF}. All equivalent width (EW) values are given in the rest frame. 

\section{Data: A2744 viewed by MUSE and \textit{JWST}/NIRCam}
\label{sect:data}
\indent We combine MUSE's IFU spectroscopy of the A2744 galaxy cluster with the full range of imaging available, using the catalogs from the UNCOVER Treasury survey (PIs Labbé and Bezanson, JWST-GO-2561; \citet{bezanson2022UNCOVER}). This photometry is among the deepest available (median catalog $5\sigma$ depth is 29.21 and individual filter depths can be found in Table 1 of \citealt{weaver2023UNCOVERcatalogs}) and ideally suited for the detection and characterisation of high-redshift galaxies.

\subsection{MUSE LAEs: LLAMAS}
\label{sect:LAE}
\indent The observations of A2744 form part of the MUSE Lensing Cluster GTO data and form part of a data release by \citet{richard2021atlas}. The MUSE Lensing Cluster programme contains clusters specifically chosen for efficient amplification of background sources such as high-redshift LAEs. Among these clusters, A2744 has been particularly well studied (e.g. the well known ASTRODEEP and HFF-DEEPSPACE catalogues \citealt{lotz2017frontier,shipley2018hff}). A2744 has four MUSE pointings (each $1\times1\,\mathrm{arcmin^2}$) forming a $2\times2$ mosaic with exposures of between 3.5 and 7 hours. Sources are identified in the MUSE datacube using the MUSELET software \citep{piqueras2017mpdaf} \footnote{\url{https://mpdaf.readthedocs.io/en/latest/muselet.html}}, the {\tt Marz} software \citep{hinton2016marz}, and the Source Inspector package \citep{bacon2022musedatareleaseII}. These packages allow the user to define redshifts for objects based on identification of emission and absorption features, as well as HST images of sources and redshift suggestions from {\tt Marz}. The spectroscopic catalog for A2744, originally presented in \citet{mahler2018strong}, improved by \citet{richard2021atlas}, is publicly available \footnote{\url{https://cral-perso.univ-lyon1.fr/labo/perso/johan.richard/MUSE_data_release/}}.\\
\indent The LAEs identified behind A2744 form part of the Lensing Lyman Alpha MUSE Arcs Sample (LLAMAS) \citep{AC2022LLAMAS}. Only LAEs with a high redshift-confidence level from the LLAMA sample are used in this work. Magnifications for all sources are noise-weighted means across each pixel in the source, calculated using the lens model from \citet{richard2021atlas} and the LENSTOOL software \citep{kneib1996lenstooloriginal,jullo2007lenstool}. For A2744, there are 154 images of 121 individual LAEs. It contains the most LAEs of all the LLAMAS clusters, making it ideal for this single cluster study. \\
\indent The flux for these LAEs is determined using the method detailed in \cite{GdlV2019LAELF} and \ttpaper. This method uses SExtractor \citep{EB96SEx} on the continuum-subtracted Narrow Band (NB) datacube in which each LAE was detected.

\subsection{\textit{JWST}/NIRCam data}
\label{sect:JWSTdata}
\begin{figure}
    \centering
    \includegraphics[width=0.5\textwidth]{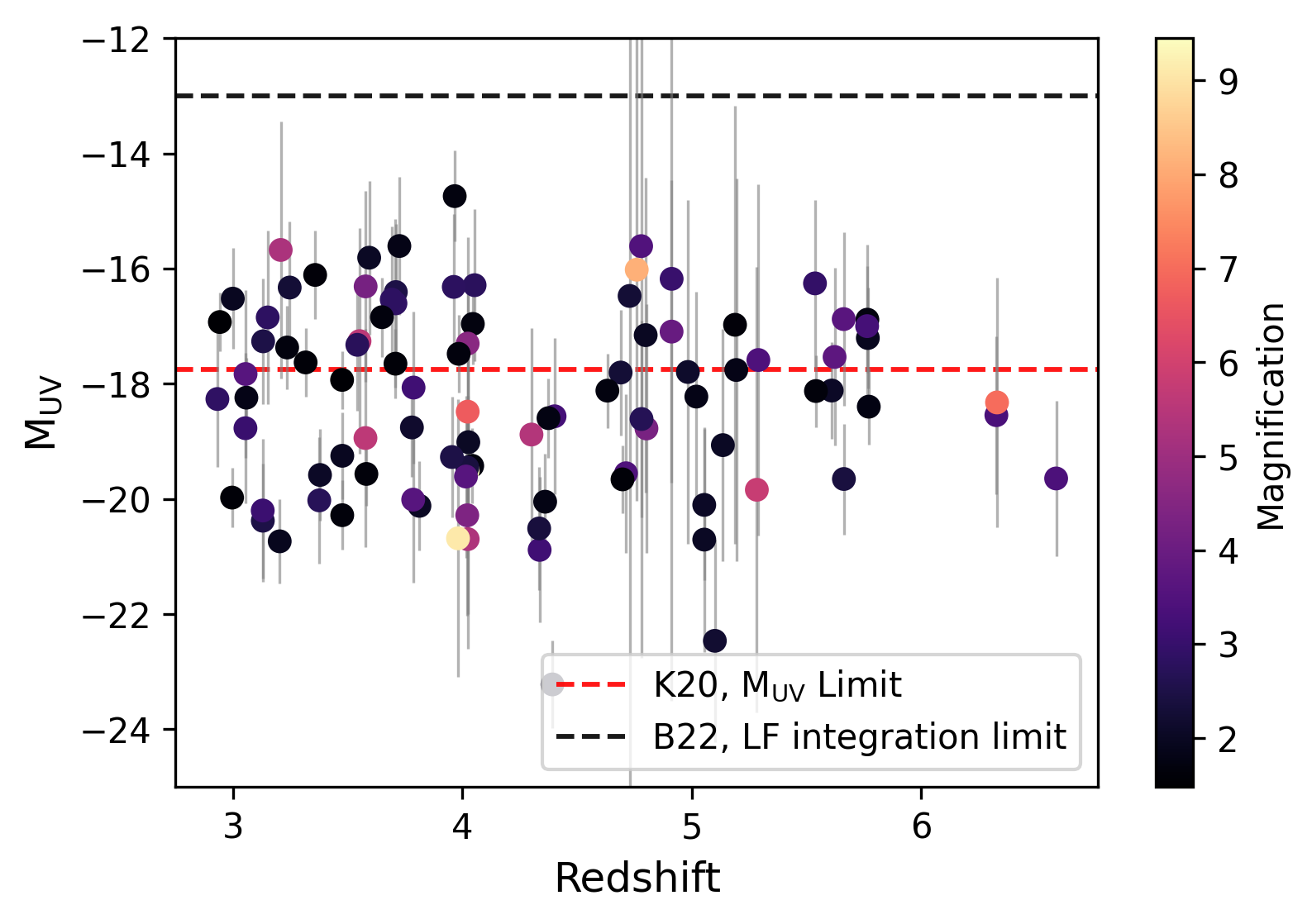}
    \caption{The distribution of the 96 individual LAEs used in this study in redshift and absolute UV magnitude, colour-coded by lensing magnification. Redshifts quoted are spectroscopic, from MUSE, and magnitude values are derived from the filter that covers the rest frame $1500\,\AA$ emission. Horizontal dashed lines denote relevant limits: red denotes the $\mathrm{M_{UV}}$ limit in an equivalent blank field study \citep{kusakabe2020} and black denotes the integration limit used when calculating the star foramtion rate density from the UV luminosity function (LF) in the latest study \citep{bouwens2022LF_II:LFs}.}
    \label{fig:z_MUV}
\end{figure}

The photometry used for this study comes from the UNCOVER treasury survey \citep{bezanson2022UNCOVER}. The photometric catalogs themselves are described in \cite{weaver2023UNCOVERcatalogs,Kokorev2022_A2744HSTdata}\footnote{\url{https://jwst-uncover.github.io}}. This survey combines HST data from the F435W, F606W, F814W, F105W, F115W, F125W, F140W and F160W filters with NIRCam data from the F090W, F150W, F200W, F277W, F356W, F410M and F444W bands. This gives excellent coverage from 15 bands between $400\,\mathrm{nm}$ and $4.5\,\mathrm{\mu m}$. \\
\indent \cite{weaver2023UNCOVERcatalogs} subtract the bright cluster galaxies (BCGs) and intra-cluster light (ICL) in order to better detect faint galaxies. The detection is then performed on a combination of the F277W, F356W, and F444W bands, PSF (point spread function)-matched to the F444W band. Two photometric catalogs are created with  a $0.32''$-aperture catalog being optimised for faint, compact sources. \\
\indent Of the 154 images of MUSE LAEs in the MUSE field of view, 121 images are detected in the UNCOVER photometry, of 99 individual LAEs. Due to the extended nature of some lensed LAEs, they are identified as two continuum objects. In this case, we combine the fluxes of the continuum objects. This happens in seven cases. In two cases, we reject the continuum associated to an LAE as it is clearly coming from a BCG. Ultimately, we chose a representative for each multiple image system with the least contamination possible. We remove one more object as it is extremely highly magnified (magnification factor 135), meaning that the uncertainty on the derived SFR and stellar mass would be too great. 96 individual LAEs remain in our sample.\\
\indent With a magnification factor range of 1.5 to 9.5 and a sample average magnification of 3, the resultant \lya luminosities probed are in the range $40<\mathrm{log_{10}(L_{Ly\alpha}/erg\,s^{-1})}<43$. The UV-magnitude range is $-23\lesssim\mathrm{M_{UV}}\lesssim-14$, where this value is derived from the filter closest to the $1500\,\AA$ rest frame emission, in the same way as \citet{goovaerts2023}. The redshift and absolute magnitude distribution is displayed for the 96 LAEs in Fig.~\ref{fig:z_MUV}, colour-coded by the lensing magnification of each galaxy. One can see from this plot that our sample populates a large part of the $\mathrm{M_{UV}}$ space between a current blank field limit for a similar study of LAEs in the literature \citep{kusakabe2020} and the typical integration limit applied to the UV luminosity function when calculating the total star formation rate density: the operative quantity when considering whether a population can reionise the neutral IGM \citep{bouwens2022LF_II:LFs}.   

\section{SED fitting with \cg}
\label{sect:cigale}
\cg (Code Investigating GALaxy Emission) \citep{boquien2019cigale} is a {\tt python} code\footnote{\url{https://cigale.lam.fr}} designed to extract galaxy properties from  photometric and spectroscopic observations ranging between the far-UV to radio. It takes into account flexible star formation histories (SFHs), nebular emission and different dust attenuation models. We fit the observed photometry of all the JWST-detected LAE images in A2744 (121 out of 154). We take advantage of the accurately known, secure spectroscopic redshifts available for these LAEs, fixing these values in {\tt CIGALE}. We also subtract \lya fluxes from the filters that see them so they do not contaminate the photometry.\\
\indent We investigate two different star-formation histories (SFHs), a single exponentially decaying burst and a double burst model with an initial burst of star formation followed by a second, delayed burst \citep{malek2018cigaleSFH,boquien2019cigale}. A range of ages are taken into account in both models, from $10-700\,\mathrm{Myr}$. Mass fractions for the delayed burst range from 0.001 to 0.65. The best-fit of these two SFHs is carried forward for analysis. \\
\indent We use the stellar library of \citet{BandC03SED} with five metallicities ranging between 0.0001 and 0.02. We take into account nebular emission as some high-redshift galaxies can exhibit strong nebular lines \citep{Maseda2020IRAC_Halpha_contam,Schaerer2022JWST_high_z_galaxies,Matthee2023JWST_nebular_lines,brinchmann2023high_z_JWST_lines}. The nebular models used by \cg (Theulé et al. A\&A, in revision) are computed using the {\tt CLOUDY} software \cite{Ferland2017CLOUDY}. Gas metallicities used range between 0.0004 and 0.02. Dust attenuation in our galaxies is modelled based on the modified starburst attenuation curve from \cite{Calzetti00dust}.\\ 
\indent \cg outputs the required physical parameters of each galaxy, in this case stellar mass, SFR, and the $\chi^2_{\nu}$ statistic for each fit. In order to check the consistency of these results and the reliability of the \cg fitting, we use \cg's {\tt mock\_flag} function to generate mock results with simulated noise, to which we compare the actual results in order to assess how reliably we estimate parameters such as the mass and SFR. This comparison is detailed in Appendix~\ref{sect:app_cg}.

\section{Results}
\label{sect:results}

\subsection{\lya Luminosity to Stellar Mass Relation}
\begin{figure*}
    \centering
    \includegraphics[width=\textwidth]{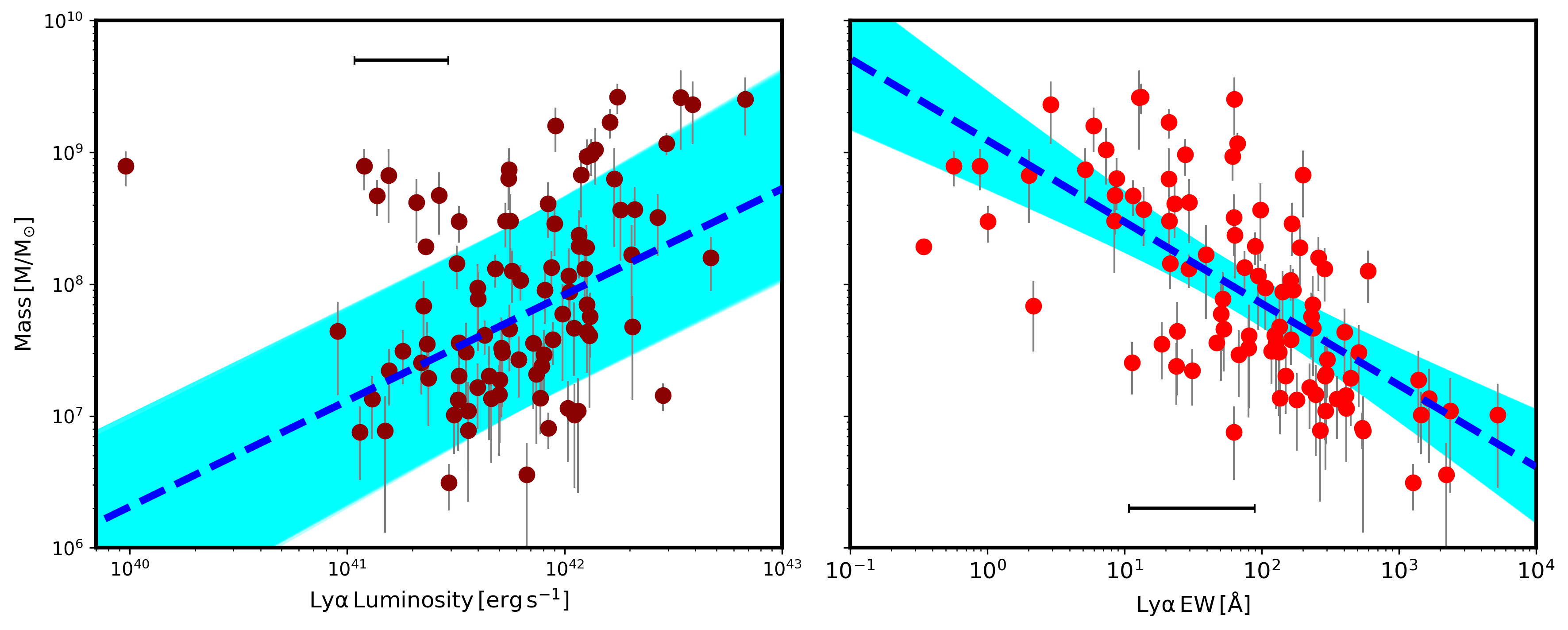}
    \caption{Stellar mass vs \lya parameters. \textit{Left:} stellar mass plotted against \lya luminosity. \textit{Right:} stellar mass plotted against \ewlya. Blue dashed lines indicate linear fits to the data and cyan shaded areas the $1\sigma$ uncertainties on the fits, details of which are given in the text. Median uncertainties on the \lya parameters are shown by the horizontal black error bars.}
    \label{fig:mass_vs_lya}
\end{figure*}
Following analysis from \cite{santos2020SC4K_LAE_MS,santos2021LAEmassrelation} we can assess the relationship between stellar mass and both \lya luminosity and \lya equivalent width (\ewlya). This can help to evaluate the connection between the host galaxy and its \lya emission. Due to the complicated nature of \lya transfer and escape \citep{verhamme2008lyadustsim,leclercq2017lyahaloes,Matthee2022what_makes_LAE,Blaizot2023lya_line}, this connection is still under much discussion. These relationships between stellar mass and \lya parameters are shown in Fig.~\ref{fig:mass_vs_lya}.\\
\indent Fitting a linear relationship between \lya luminosity and stellar mass for our 96 LAEs, we find $\mathrm{log(M_{\star}/M_{\odot})=(0.85\pm0.17)\, log(L_{Ly\alpha} /erg s^{−1}) - (27.8\pm6.4)}$. This result has a large dispersion but is in rough agreement with that in \cite{santos2021LAEmassrelation} (albeit with a lower slope and normalisation), which is especially interesting as the 4000 SC4K LAEs in that study are in a \lya luminosity range of $42.5\lesssim \mathrm{log(L_{Ly\alpha}) (erg\,s^{-1})}\lesssim44$, therefore far brighter than our sample. Our results suggest that this relationship holds down to intrinsically faint LAEs: when a galaxy displays \lya emission, the connection between the stellar mass of the galaxies and the amount of \lya luminosity they emit remains roughly constant across all currently-accessible luminosities. However, there is significant scatter in the data, which is to be expected, and indicative of the wide variety of conditions within galaxies affecting \lya escape. If a given galaxy lies far above this best-fit relation (blue dashed line in Fig.~\ref{fig:mass_vs_lya}), it is emitting far less \lya emission than expected for a galaxy of its mass. This likely indicates other factors at play, such as higher dust content or specific viewing angles \citep{Atek2008lya_dust,verhamme2008lyadustsim,Chen2023Lya_escape_JWST}. Given the dispersion in the left panel of Fig.~\ref{fig:mass_vs_lya}, these are clearly important effects. For example, the object at $10^{40}\,\mathrm{erg\,s^{-1}}$ lies $\sim3\sigma$ away from the best fit relation, indicating that some property of this galaxy is preventing the escape of \lya photons more than usual for LAEs at this redshift and mass-range. Estimating dust content and extinction from longer wavelength data could help to shed more light on this issue. \\
\indent When comparing \ewlya with stellar mass, we find a linear relationship of the form: $\mathrm{log(M_{\star}/M_{\odot})=(- 0.62\pm0.07)\, log(EW_{Ly\alpha}) +(9.1\pm0.1)}$. This is a tighter relationship than for luminosity, which is unsurprising as the UV magnitude used to calculate \ewlya is tightly correlated with the stellar mass \citep{duncan2014masslight,santini2023masslightGLASS}. This suggests that \lya photons find it relatively harder to escape higher mass galaxies. This is likely due to the higher dust content in galaxies of higher mass \citep{zahid2013stellar_mass_dust,Calura2017dust_stellar_mass,Donevski2020dust_stellar_mass}. While this relation is tighter than that between stellar mass and luminosity, a dispersion is still evident around the best-fit relation. This indicates that for a given mass, a range of \lya escape scenarios are possible. A more complete study on these relations necessitates larger samples of lensed galaxies which can be binned in terms of stellar mass. We return to this in Sect.~\ref{sect:conclusion}.  
 
\subsection{Star-Formation Main Sequence}
\label{sect:MS}


\begin{figure*}
    \centering
    \includegraphics[width=0.85\textwidth]{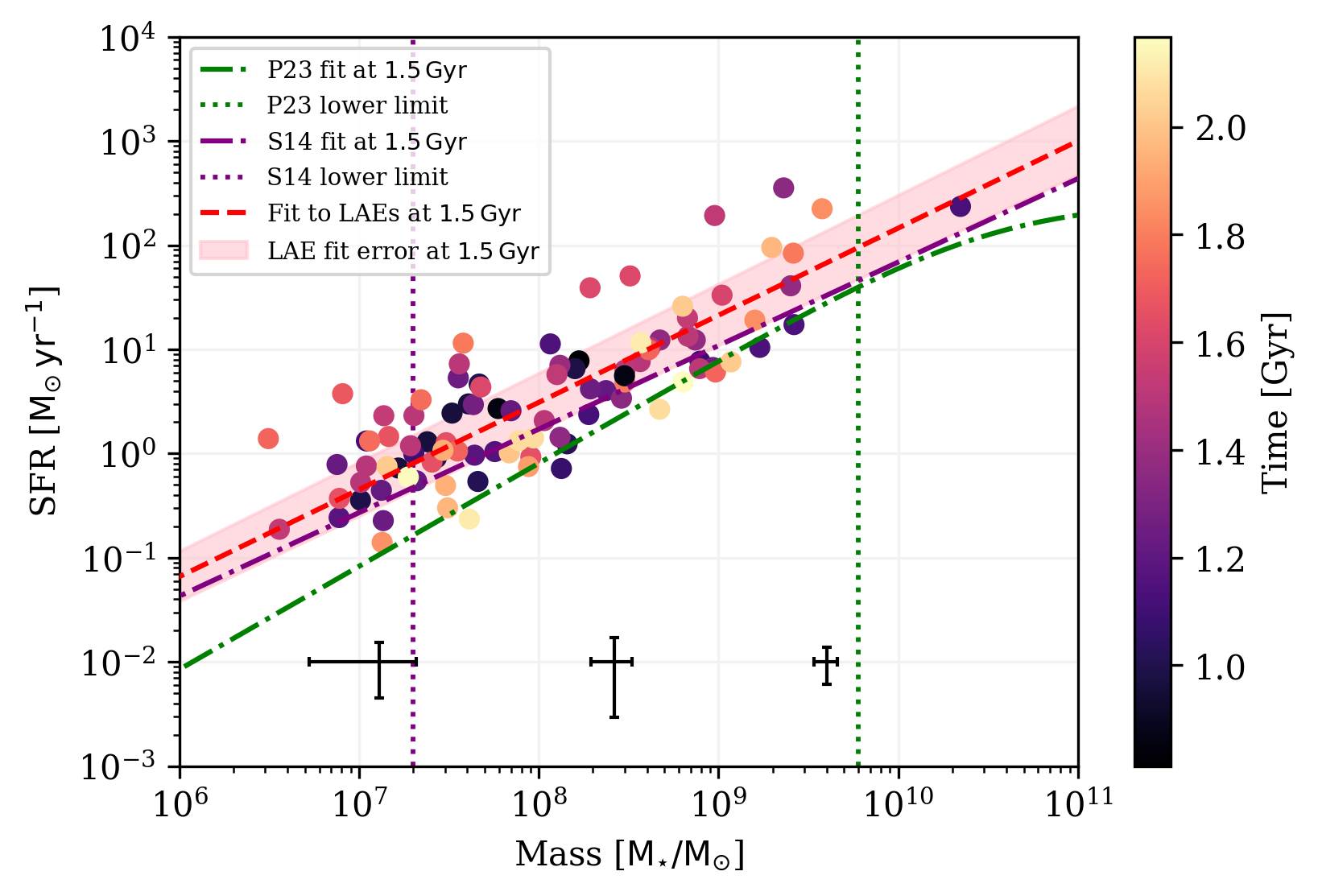}
    \\[\smallskipamount]
    \includegraphics[width=0.85\textwidth]{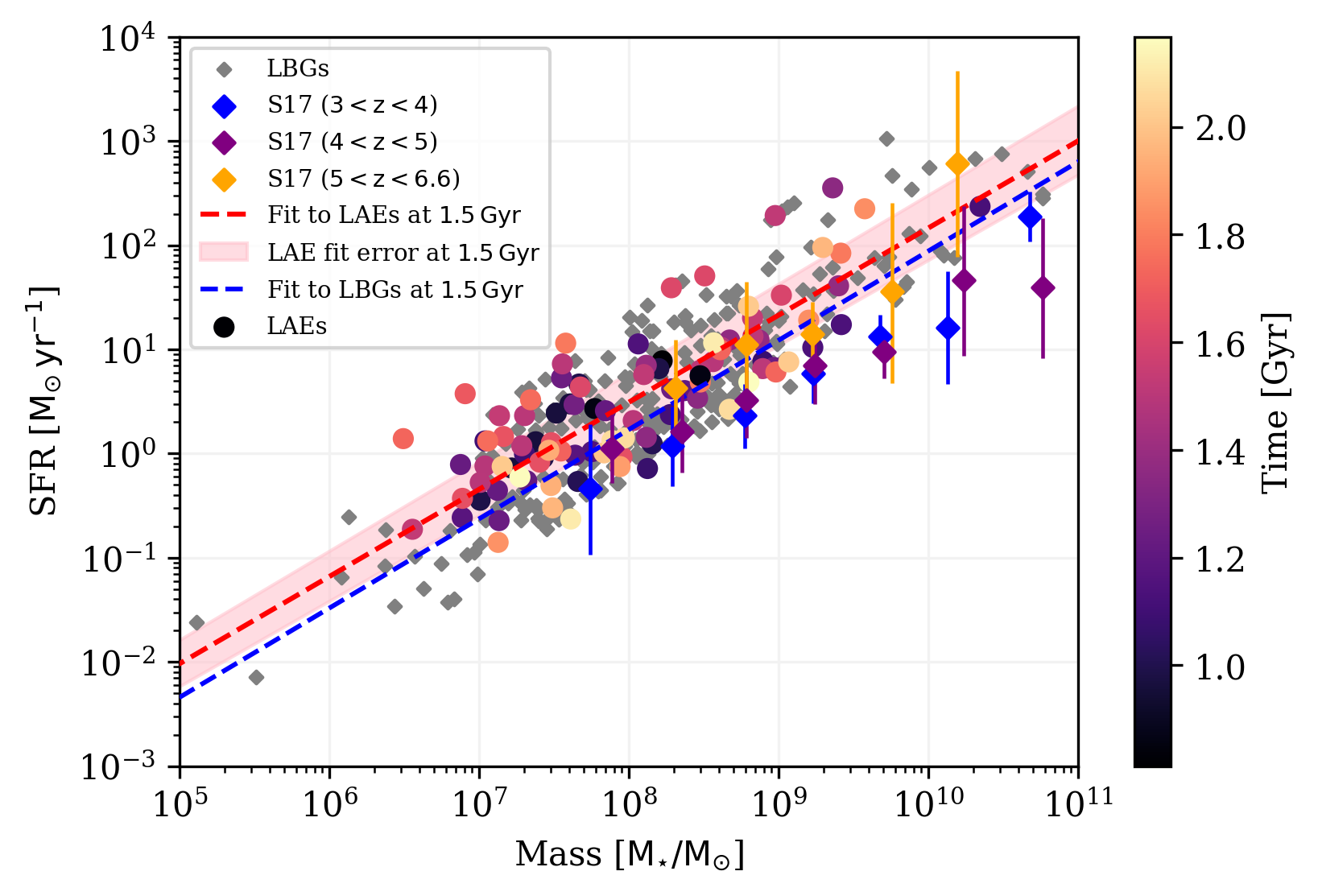}
    \caption{Stellar mass vs SFR: the galaxy MS.\\
    \textit{Upper:} The 96 LAEs in our final sample plotted on the star-formation MS, together with the best fits to the data of \citetalias{P23MS}, \citetalias{S14MS} and the best fit to the LAEs following Eq.~\ref{eq:SF_speagle}. LAEs are colour-coded by ``Time'', in Gyr, which is the age of the Universe at the observed redshift. The purple, dotted line indicates the lowest-stellar mass limit of \citetalias{S14MS}'s review ($2\times10^7\,\mathrm{M_{\odot}}$) and the green, dotted line indicates the lowest-stellar mass limit of \citetalias{P23MS}'s review ($6\times10^9\,\mathrm{M_{\odot}}$). The pink shaded area denotes the $\pm1\sigma$ area for the best-fit MS relation at the sample mean age, $\sim\mathrm{1.48\,Gyr}$. Mean error bars in three mass bins ($10^6<\mathrm{M/M_{\odot}}<2.5\times10^7$, $2.5\times10^7<\mathrm{M_{\star}/M_{\odot}}<5\times10^8$ and $5\times10^8<\mathrm{M_{\star}/M_{\odot}}<10^{10}$) are shown in black at the bottom of the plot.\\
    \textit{Lower:} The same LAEs and best-fit relation, compared this time to results in the same mass range but UV-selected: the LBGs from A2744 from \igpaper as grey diamonds, with a blue best-fit relation, and the mass-binned results from \citetalias{S17MS} as coloured diamonds with error bars.
    }    
    \label{fig:MS}
\end{figure*}

We plot the best-fit stellar masses and SFRs derived from \cg in Fig.~\ref{fig:MS}. This is done for the first time for lensed LAEs of these masses at these redshifts. Therefore, in order to gain an understanding of how these galaxies compare to galaxies of higher mass and UV-selected samples, we incorporate the appropriate best fits from the reviews of \citetalias{S14MS} and \citetalias{P23MS}, as well as data from \citetalias{S17MS}, a similar study to ours (without the benefit of JWST data) using Lyman break-selected galaxies (LBGs) in the Hubble Frontier Fields (HFF). Additionally, we compare the 96 LAEs to the sample of LBGs selected behind A2744 from \igpaper (specifically, the LBG-only galaxies, which show no \lya emission). These have undergone an identical fitting process to the LAEs, detailed in the previous section. We chose to adopt the time-dependent form of \citetalias{S14MS} to fit to our data. As we do not probe the higher-mass end of the MS, where its linearity breaks down (see \citetalias{P23MS} and references therein), this linear form for the MS is safe for us to adopt. This form of the MS is given as: \\
\begin{equation}
    \label{eq:SF_speagle}
    \mathrm{log\,SFR}=(a - b\times t)\,\mathrm{log\,M_{\star}} - ( c - d\times t)
\end{equation}
The values of the constants $a, b, c,$ and $ d$ are given in Table~\ref{tab:MS_results}, both for this work and for those we compare to. We adjust the relations quoted in \citetalias{S14MS} and \citetalias{P23MS} to take into account the different IMFs used. We use the Salpeter IMF \citep{salpeter1955IMF} in our SED fitting but the Kroupa IMF \citep{Kroupa2001IMF} is used in both of the reviews we compare to. This necessitates a correction to stellar masses which we make following the prescription in both \citetalias{S14MS} and \citetalias{P23MS} (taken from \citealt{Zahid2012IMFcorrections}) to $\mathrm{M_{\star,K}=0.62M_{\star,S}}$ where $\mathrm{M_{\star,K}}$ and $\mathrm{M_{\star,K}}$ are the stellar masses calculated using a Kroupa IMF and a Salpeter IMF, respectively. These corrections are incorporated in the dashed lines in Fig.~\ref{fig:MS} but the original quoted fit values are given in Table~\ref{tab:MS_results}.\\
\indent It is immediately clear from the upper panel of Fig.~\ref{fig:MS} that the low-mass LAEs studied here lie above the fits to the MS calibrated at higher masses. The question is, then, why do these low-mass LAEs lie above these MS relations? Is this a result of the selection of these galaxies as LAEs, or is this rather a result of their low mass? We seek to answer this through comparison to the aforementioned datasets. The most stark difference is seen when comparing to the fit of \citetalias{P23MS}. The authors of this review fit their data to a relation of the form:
\begin{equation}
    \mathrm{log\,SFR=}\ a_0+a_1t-\mathrm{log(1+(M_{\star}}/10^{a_2+a_3t})^{-a_4}) 
\end{equation}
with $a_0 = 2.693\pm0.012$, $a_1=-0.186\pm0.009$, $a_2=10.85\pm0.05$, $a_3=-0.0729\pm0.0024$ and $a_4=0.99\pm0.01$. This fit is designed to take into account the turnover of the MS at high masses ($\mathrm{M_{\star}}>10^{10}$), however, it performs poorly when compared to our sample, especially in the low-mass regime. We see a 0.43 dex difference at $10^7\,\mathrm{M_{\odot}}$ and a 0.19 dex difference at $10^{10}\,\mathrm{M_{\odot}}$. This difficulty in fitting the low-mass LAEs is unsurprising as the \citetalias{P23MS} relation is derived using galaxies down to a lower mass limit of $10^{8.5}\,\mathrm{M_{\odot}}$.\\
\indent The fit of \citetalias{S14MS}, of the form shown in Eq.~\ref{eq:SF_speagle}, fits the data better, noting the error of $\sim0.3$ dex in their MS relation. We see a 0.13 dex difference at $10^7\,\mathrm{M_{\odot}}$ and a 0.17 dex difference at $10^{10}\,\mathrm{M_{\odot}}$, with \citetalias{S14MS}'s relation having been derived using galaxies of mass down to $10^{7.3}\,\mathrm{M_{\odot}}$. However only two studies with galaxies of mass below $10^{8}\,\mathrm{M_{\odot}}$ are included, and \citetalias{S14MS} do not incorporate the first two Gyr of galaxy evolution in their fit, limiting usefulness as a comparative tool to this study. Nevertheless, this best-fit relation is consistent with ours to within their respective errors. \\
\indent The fit of the form in Eq.~\ref{eq:SF_speagle} that we find for our LAEs is $\mathrm{log\,SFR}=(0.88\pm0.07 - 0.030\pm0.027\times t)\,\mathrm{log\,M_{\star}} - ( 6.31\pm0.41 - 0.08\pm0.37\times t)$. The $\pm1\sigma$ error region is shaded in pink on both plots in Fig.~\ref{fig:MS}. We stress that, given the conclusions above, this relation and uncertainty likely does not hold for the lower (and higher)-mass regions of this graph, which our data does not probe.\\
\indent The non-time-dependent constants of our best fit ($a$ and $c$ in Eq.~\ref{eq:SF_speagle}), are consistent within their uncertainties with the best-fit of \citetalias{S14MS}, and we see a similar but less significant evolution with time (constants $b$ and $d$ in Eq.~\ref{eq:SF_speagle}). This points to galaxies selected as LAEs across our redshift range as being consistently young and highly star-forming. However, a less significant evolution may also be due in part to our sample being smaller than that of \citetalias{S14MS}. Details of the fitting for the LAEs and LBGs in this work, as well as comparisons to \citetalias{S14MS} and \cite{S17MS} (henceforth \citetalias{S17MS}) are given in Table~\ref{tab:MS_results}.\\
\indent The discrepancy between our results and the collations of \citetalias{S14MS} and \citetalias{P23MS}, as mentioned, likely comes from the difference in mass-range probed, but to further verify this, we compare to two UV selected samples in the lower plot of Fig.~\ref{fig:MS}: \citetalias{S17MS}, conducted in the HFF galaxy clusters, and the selection of LBGs from \igpaper, (294 galaxies), selected behind A2744 itself, in the same volume as the LAEs in this work. The fit to the LBGs (blue dashed line) is $(0.89\pm0.11 - 0.024\pm0.007\times t)\,\mathrm{logM_{\star}} - ( 6.75\pm0.78 - 0.09\pm0.59\times t)$. This relation is consistent with that for the LAEs, further suggesting that it is the mass (and redshift) range probed, rather than the selection of the sample by \lya emission that differentiates these galaxies from MS relations in the literature. A small effect can be seen in the lower panel of Fig.~\ref{fig:MS}; the LAE and LBG best fits agree better in the higher-mass range than the lower mass-range, but this effect is at low confidence ($<1\sigma$) for this dataset.\\ 
\indent The mass-binned data of \citetalias{S17MS} are in good agreement with our data, particularly in the highest redshift bin ($5<z<6.6$). They also find a similar trend when considering earlier times (higher redshifts): an increasing normalisation but a similar slope compared to later times (lower redshifts). However this evolution is more dramatic in their data than ours. This could be due to their larger sample size, however uncertainties related to the lack of wavelength coverage past $400\,\mathrm{nm}$ above $z=3$ may have an effect. This photometry can be especially problematic in crowded fields such as lensing clusters \citep{Laporte2015spitzerlensing,goovaerts2023}. The $2\sigma$-clipping procedure used in \citetalias{S17MS} and the choice to calculate the MS according to $\mathrm{log\,SFR = \alpha\, log\,(M/M_{9.7})+\beta}$ (where $M_{9.7}=10^{9.7}\,\mathrm{M_{\odot}}$), may also play a role.\\
\indent The role of stellar mass is additionally supported by very recent results from \cite{Nakane2023highzLAEs}, who find that high-redshift ($z>7$) LAEs (and LBGs) in a higher mass range than this study (most galaxies have $\mathrm{M_{\star}/M_{\odot}>10^8}$) typically lie on the MS (the reference MS in this case is from \citetalias{S17MS}), although these results, like ours, see a lot of scatter in individual data points. The authors, similarly to this study, make no significant distinction between the LAE and non-LAE populations in terms of their position on the MS. \\
\indent Viewed together, these results indicate that, while the low-mass LAEs in this study reside above the typical MS estimates derived from higher-mass samples, particularly that of \citetalias{P23MS}, derived for galaxies of much higher mass, they are likely similar to galaxies that are not LAEs, but are of similar mass at similar redshifts. This points to an enhanced specific SFR in these low mass populations, but no special role for LAEs themselves. In the context of cosmic reionisation, this supports conclusions that low-mass galaxies play an important role, however LAEs are not singular in their impact on this process.

\begin{table*}[]
    \centering
    \caption{Results for the galaxy MS from the literature and from this study.}
    \begin{tabular}{c|c|c|c|c|c}
        \hline
        \hline
         Study & Redshift & $a$ & $b$ & $c$ & $d$ \\
         \hline
         \citetalias{S14MS} & $0<z<6$ & $0.84\pm0.02$ & $-0.026\pm0.003$ & $6.51\pm0.24$ & $-0.11\pm0.03$\\
         \citetalias{S17MS} & $3<z<4$ & $1.02\pm0.04$ & -- & $8.52\pm0.42$ & -- \\
         \citetalias{S17MS} & $4<z<5$ & $0.94\pm0.06$ & -- & $7.75\pm0.63$ & -- \\
         \citetalias{S17MS} & $5<z<6.6$ & $0.92\pm0.15$ & -- & $ 6.93\pm1.59$ & -- \\
         \textbf{This work: LAEs} & $2.9<z<6.7$ & $0.88\pm0.06$ & $-0.030\pm0.027$ & $6.31\pm0.41$ & $-0.08\pm0.37$ \\
         \textbf{This work: LBGs} & $2.9<z<6.7$ & $0.89\pm0.11$ & $-0.024\pm0.007$ & $6.75\pm0.78$ & $-0.09\pm0.59$\\
         \hline
         \hline
    \end{tabular}
    \vspace{1mm}
    \begin{tablenotes}
      \small
      \item \textbf{Notes.} Constants $a$, $b$, $c$ and $d$ refer to the constants from Eq.~\ref{eq:SF_speagle}. For \citetalias{S17MS} a time dependance (constants $b$ and $d$) is not quoted so have been omitted. 
    \end{tablenotes}
    \label{tab:MS_results}
\end{table*}

\section{Conclusions}
\label{sect:conclusion}

With new \textit{JWST}/NIRCam data, secure spectroscopic redshifts from VLT/MUSE and \cg SED fitting, we have obtained reliable estimates for the SFR and stellar mass for intrinsically-faint, lensed, reionisation-era LAEs. This allowed us to place these galaxies on the star-formation main sequence and compare them to UV-selected samples as well as studies conducted at higher mass. We also compared stellar masses for these galaxies with \lya observables.\\
\indent Our conclusions are summarised thus:\\
\indent -- The relation we find for stellar mass and \lya luminosity: $\mathrm{log(M_{\star}/M_{\odot})=(0.85\pm0.17)\, log(L_{Ly\alpha} /erg s^{−1}) - (27.8\pm6.4)}$, agrees with similar relations found for samples of LAEs with higher masses. This indicates that the general relation between stellar mass and the amount of \lya emission observed from a galaxy holds well over the range of currently observed LAEs.\\
\indent -- We note that this relation has a large dispersion, reflecting the scatter of the data. This indicates that there is a range of \lya escape scenarios for galaxies of the same mass. This likely depends on the dust extinction of the galaxies, as well as viewing angle and geometry, but longer-wavelength observations quantifying dust content and extinction are necessary to confirm this. \\
\indent -- The relation between \lya EW and stellar mass is tighter than for \lya luminosity: $\mathrm{log(M_{\star}/M_{\odot})=(- 0.62\pm0.07)\, log(EW_{Ly\alpha}) +(9.1\pm0.1)}$. This anticorrelation suggests that \lya photons find it harder to escape higher-mass galaxies. \\
\indent -- The MS relation we find for our LAEs between redshifts of 2.9 and 6.7, and masses of $10^6\,\mathrm{M_{\odot}}$ and $10^{10}\,\mathrm{M_{\odot}}$ is $\mathrm{log\,SFR}=(0.88\pm0.07 - 0.030\pm0.027\times t)\,\mathrm{log\,M_{\star}} - ( 6.31\pm0.41 - 0.08\pm0.37\times t)$. This relation is consistent with the best-fit found in \citetalias{S14MS}, which is derived using galaxies down to $\mathrm{10^{7.3}\,M_{\odot}}$, and agrees well with galaxies not selected as LAEs at similar masses and redshifts. However a clear discrepancy is seen between this relation and the best-fit of \citetalias{P23MS} which is derived only using galaxies of stellar mass greater than $\mathrm{10^{8.5}\,M_{\odot}}$.\\
\indent -- Our results suggest that the mass range probed is the main influence in our determination of the galaxy main sequence relation, more than the selection of our sample by their \lya emission. When considering the context of cosmic reionisation, while low-mass galaxies have been shown to play a significant role, which our results support, our results also suggest that LAEs do not distinguish themselves in this process as they have no particular enhanced sSFR compared to the general population of star-forming galaxies at this time.\\
\indent Larger samples of this type, for more precise estimations of the main sequence relation, both for LAEs and UV-selected samples, will have to wait until further imaging of lensing clusters with \textit{JWST}/NIRCam. Observing many lensing clusters with the required depth of photometry is expensive, and so far A2744 has been prioritised. However in the coming months and years, we can expect samples of this type to grow significantly, with the addition of further lensing clusters containing hundreds of LAEs in and around the epoch of reionisation. Samples with \textit{JWST}/NIRSpec data are doubly valuable, enabling SFR determinations using H$\alpha$ luminosities for these redshifts, as well as full spectral fitting for more accurate galaxy properties. Large samples with accurate H$\alpha$ luminosity determinations are hard to obtain, and have not been a priority for the redshifts concerned by this work. Nevertheless, in future JWST Cycles larger samples with better data are likely to become available.


\begin{acknowledgements}
The authors would like to thank the anonymous referee for numerous useful comments which helped in improving this article.
This work is done based on observations made with ESO Telescopes
at the La Silla Paranal Observatory under programme IDs 060.A-9345, 092.A-0472,
094.A-0115, 095.A-0181, 096.A-0710, 097.A0269, 100.A-0249, and
294.A-5032. Also based on observations obtained with the NASA/ESA Hubble
Space Telescope, retrieved from the Mikulski Archive for Space Telescopes
(MAST) at the Space Telescope Science Institute (STScI). STScI is operated by
the Association of Universities for Research in Astronomy, Inc. under NASA
contract NAS 5-26555. 
All plots in this paper were created using Matplotlib (Hunter 2007). 
Part of this work was supported by the French CNRS, the Aix-Marseille University, the French Programme National de Cosmologie et Galaxies (PNCG) of CNRS/INSU with INP and IN2P3, co-funded by CEA and CNES.
This work also received support from the French government under the France 2030 investment plan, as part of the Excellence Initiative of Aix-Marseille University - A*MIDEX (AMX-19-IET-008 - IPhU).
Financial support from the World Laboratory, the Odon Vallet Foundation and VNSC is gratefully acknowledged. Tran Thi Thai was funded by Vingroup JSC and supported by the Master, PhD Scholarship Programme of Vingroup Innovation Foundation (VINIF), Institute of Big Data, code VINIF.2023.TS.108. Pham Tuan-Anh was funded by Vingroup Innovation Foundation (VINIF) under project code VINIF.2023.DA.057.       
\end{acknowledgements}

%
%
%

\bibliographystyle{aa} 
\bibliography{aanda.bib} 

\begin{thebibliography}{90}
\expandafter\ifx\csname natexlab\endcsname\relax\def\natexlab#1{#1}\fi

\bibitem[{Arrabal~Haro {et~al.}(2018)Arrabal~Haro, Rodr{\'\i}guez~Espinosa, Mu{\~n}oz-Tu{\~n}{\'o}n, P{\'e}rez-Gonz{\'a}lez, Dannerbauer, Bongiovanni, Barro, Cava, Lumbreras-Calle, Hern{\'a}n-Caballero, {et~al.}}]{arrabal2018LAEfrac}
Arrabal~Haro, P., Rodr{\'\i}guez~Espinosa, J., Mu{\~n}oz-Tu{\~n}{\'o}n, C., {et~al.} 2018, Monthly Notices of the Royal Astronomical Society, 478, 3740

\bibitem[{{Atek} {et~al.}(2008){Atek}, {Kunth}, {Hayes}, {{\"O}stlin}, \& {Mas-Hesse}}]{Atek2008lya_dust}
{Atek}, H., {Kunth}, D., {Hayes}, M., {{\"O}stlin}, G., \& {Mas-Hesse}, J.~M. 2008, \aap, 488, 491

\bibitem[{Atek {et~al.}(2023)Atek, Labbé, Furtak, Chemerynska, Fujimoto, Setton, Miller, Oesch, Bezanson, Price, Dayal, Zitrin, Kokorev, Weaver, Brammer, van Dokkum, Williams, Cutler, Feldmann, Fudamoto, Greene, Leja, Maseda, Muzzin, Pan, Papovich, Nelson, Nanayakkara, Stark, Stefanon, Suess, Wang, \& Whitaker}]{Atek2023arxiv_lowmassgalaxies}
Atek, H., Labbé, I., Furtak, L.~J., {et~al.} 2023 [\eprint{Arxiv:2308.08540v1}]

\bibitem[{Bacon {et~al.}(2022)Bacon, Brinchmann, Conseil, Maseda, Nanayakkara, Wendt, Bacher, Mary, Weilbacher, Krajnovic, {et~al.}}]{bacon2022musedatareleaseII}
Bacon, R., Brinchmann, J., Conseil, S., {et~al.} 2022, arXiv preprint arXiv:2211.08493

\bibitem[{Bertin \& Arnouts(1996)}]{EB96SEx}
Bertin, E. \& Arnouts, S. 1996, Astronomy and astrophysics supplement series, 117, 393

\bibitem[{{Bezanson} {et~al.}(2022){Bezanson}, {Labbe}, {Whitaker}, {Leja}, {Price}, {Franx}, {Brammer}, {Marchesini}, {Zitrin}, {Wang}, {Weaver}, {Furtak}, {Atek}, {Coe}, {Cutler}, {Dayal}, {van Dokkum}, {Feldmann}, {Forster Schreiber}, {Fujimoto}, {Geha}, {Glazebrook}, {de Graaff}, {Greene}, {Juneau}, {Kassin}, {Kriek}, {Khullar}, {Maseda}, {Mowla}, {Muzzin}, {Nanayakkara}, {Nelson}, {Oesch}, {Pacifici}, {Pan}, {Papovich}, {Setton}, {Shapley}, {Smit}, {Stefanon}, {Taylor}, \& {Williams}}]{bezanson2022UNCOVER}
{Bezanson}, R., {Labbe}, I., {Whitaker}, K.~E., {et~al.} 2022, arXiv e-prints, arXiv:2212.04026

\bibitem[{{Blaizot} {et~al.}(2023){Blaizot}, {Garel}, {Verhamme}, {Katz}, {Kimm}, {Michel-Dansac}, {Mitchell}, {Rosdahl}, \& {Trebitsch}}]{Blaizot2023lya_line}
{Blaizot}, J., {Garel}, T., {Verhamme}, A., {et~al.} 2023, \mnras, 523, 3749

\bibitem[{Bolan {et~al.}(2022)Bolan, Lemaux, Mason, Brada{\v{c}}, Treu, Strait, Pelliccia, Pentericci, \& Malkan}]{bolan2022neutralfraction}
Bolan, P., Lemaux, B.~C., Mason, C., {et~al.} 2022, Monthly Notices of the Royal Astronomical Society, 517, 3263

\bibitem[{{Bolzonella} {et~al.}(2010){Bolzonella}, {Kova{\v{c}}}, {Pozzetti}, {Zucca}, {Cucciati}, {Lilly}, {Peng}, {Iovino}, {Zamorani}, {Vergani}, {Tasca}, {Lamareille}, {Oesch}, {Caputi}, {Kampczyk}, {Bardelli}, {Maier}, {Abbas}, {Knobel}, {Scodeggio}, {Carollo}, {Contini}, {Kneib}, {Le F{\`e}vre}, {Mainieri}, {Renzini}, {Bongiorno}, {Coppa}, {de la Torre}, {de Ravel}, {Franzetti}, {Garilli}, {Le Borgne}, {Le Brun}, {Mignoli}, {Pell{\'o}}, {Perez-Montero}, {Ricciardelli}, {Silverman}, {Tanaka}, {Tresse}, {Bottini}, {Cappi}, {Cassata}, {Cimatti}, {Guzzo}, {Koekemoer}, {Leauthaud}, {Maccagni}, {Marinoni}, {McCracken}, {Memeo}, {Meneux}, {Porciani}, {Scaramella}, {Aussel}, {Capak}, {Halliday}, {Ilbert}, {Kartaltepe}, {Salvato}, {Sanders}, {Scarlata}, {Scoville}, {Taniguchi}, \& {Thompson}}]{Bolzonella2010SEDfittingmass}
{Bolzonella}, M., {Kova{\v{c}}}, K., {Pozzetti}, L., {et~al.} 2010, \aap, 524, A76

\bibitem[{Boogaard {et~al.}(2018)Boogaard, Brinchmann, Bouch{\'e}, Paalvast, Bacon, Bouwens, Contini, Gunawardhana, Inami, Marino, {et~al.}}]{boogaard2018low_mass_MS}
Boogaard, L.~A., Brinchmann, J., Bouch{\'e}, N., {et~al.} 2018, Astronomy \& Astrophysics, 619, A27

\bibitem[{Boquien {et~al.}(2019)Boquien, Burgarella, Roehlly, Buat, Ciesla, Corre, Inoue, \& Salas}]{boquien2019cigale}
Boquien, M., Burgarella, D., Roehlly, Y., {et~al.} 2019, Astronomy \& Astrophysics, 622, A103

\bibitem[{{Bosman} {et~al.}(2022){Bosman}, {Davies}, {Becker}, {Keating}, {Davies}, {Zhu}, {Eilers}, {D'Odorico}, {Bian}, {Bischetti}, {Cristiani}, {Fan}, {Farina}, {Haehnelt}, {Hennawi}, {Kulkarni}, {Mesinger}, {Meyer}, {Onoue}, {Pallottini}, {Qin}, {Ryan-Weber}, {Schindler}, {Walter}, {Wang}, \& {Yang}}]{Bosman2022z5.3reionisation}
{Bosman}, S. E.~I., {Davies}, F.~B., {Becker}, G.~D., {et~al.} 2022, \mnras, 514, 55

\bibitem[{Bouwens {et~al.}(2022{\natexlab{a}})Bouwens, Illingworth, Ellis, Oesch, Paulino-Afonso, Ribeiro, \& Stefanon}]{bouwens2022LF_I:sample_selection}
Bouwens, R., Illingworth, G., Ellis, R., {et~al.} 2022{\natexlab{a}}, The Astrophysical Journal, 931, 81

\bibitem[{Bouwens {et~al.}(2022{\natexlab{b}})Bouwens, Illingworth, Ellis, Oesch, \& Stefanon}]{bouwens2022LF_II:LFs}
Bouwens, R., Illingworth, G., Ellis, R., Oesch, P., \& Stefanon, M. 2022{\natexlab{b}}, The Astrophysical Journal, 940, 55

\bibitem[{Bouwens {et~al.}(2021)Bouwens, Oesch, Stefanon, Illingworth, Labb{\'e}, Reddy, Atek, Montes, Naidu, Nanayakkara, {et~al.}}]{bouwens2021UVLF}
Bouwens, R., Oesch, P., Stefanon, M., {et~al.} 2021, The Astronomical Journal, 162, 47

\bibitem[{Bouwens {et~al.}(2015{\natexlab{a}})Bouwens, Illingworth, Oesch, Trenti, Labb{\'e}, Bradley, Carollo, Van~Dokkum, Gonzalez, Holwerda, {et~al.}}]{bouwens2015UVLF}
Bouwens, R.~J., Illingworth, G., Oesch, P., {et~al.} 2015{\natexlab{a}}, The Astrophysical Journal, 803, 34

\bibitem[{Bouwens {et~al.}(2015{\natexlab{b}})Bouwens, Illingworth, Oesch, Caruana, Holwerda, Smit, \& Wilkins}]{Bouwens2015planckreionisation}
Bouwens, R.~J., Illingworth, G.~D., Oesch, P.~A., {et~al.} 2015{\natexlab{b}}, The Astrophysical Journal, 811, 140

\bibitem[{Brinchmann(2023)}]{brinchmann2023high_z_JWST_lines}
Brinchmann, J. 2023, Monthly Notices of the Royal Astronomical Society, stad1704

\bibitem[{Brinchmann {et~al.}(2004)Brinchmann, Charlot, White, Tremonti, Kauffmann, Heckman, \& Brinkmann}]{brinchmann2004MSlowz}
Brinchmann, J., Charlot, S., White, S.~D., {et~al.} 2004, Monthly notices of the royal astronomical society, 351, 1151

\bibitem[{Bruzual \& Charlot(2003)}]{BandC03SED}
Bruzual, G. \& Charlot, S. 2003, Monthly Notices of the Royal Astronomical Society, 344, 1000

\bibitem[{Buat {et~al.}(2014)Buat, Heinis, Boquien, Burgarella, Charmandaris, Boissier, Boselli, Le~Borgne, \& Morrison}]{buat2014SED}
Buat, V., Heinis, S., Boquien, M., {et~al.} 2014, Astronomy \& Astrophysics, 561, A39

\bibitem[{{Calura} {et~al.}(2017){Calura}, {Pozzi}, {Cresci}, {Santini}, {Gruppioni}, {Pozzetti}, {Gilli}, {Matteucci}, \& {Maiolino}}]{Calura2017dust_stellar_mass}
{Calura}, F., {Pozzi}, F., {Cresci}, G., {et~al.} 2017, \mnras, 465, 54

\bibitem[{Calzetti {et~al.}(2000)Calzetti, Armus, Bohlin, Kinney, Koornneef, \& Storchi-Bergmann}]{Calzetti00dust}
Calzetti, D., Armus, L., Bohlin, R.~C., {et~al.} 2000, The Astrophysical Journal, 533, 682

\bibitem[{Caruana {et~al.}(2018)Caruana, Wisotzki, Herenz, Kerutt, Urrutia, Schmidt, Bouwens, Brinchmann, Cantalupo, Carollo, {et~al.}}]{caruana2018}
Caruana, J., Wisotzki, L., Herenz, E.~C., {et~al.} 2018, Monthly Notices of the Royal Astronomical Society, 473, 30

\bibitem[{Chen {et~al.}(2023)Chen, Stark, Mason, Topping, Whitler, Tang, Endsley, \& Charlot}]{Chen2023Lya_escape_JWST}
Chen, Z., Stark, D.~P., Mason, C., {et~al.} 2023 [\eprint{Arxiv:2311.13683v1}]

\bibitem[{Ciesla {et~al.}(2017)Ciesla, Elbaz, \& Fensch}]{ciesla2017sfrmass}
Ciesla, L., Elbaz, D., \& Fensch, J. 2017, Astronomy \& Astrophysics, 608, A41

\bibitem[{Claeyssens {et~al.}(2022)Claeyssens, Richard, Blaizot, Garel, Kusakabe, Bacon, Bauer, Guaita, Jeanneau, Lagattuta, {et~al.}}]{AC2022LLAMAS}
Claeyssens, A., Richard, J., Blaizot, J., {et~al.} 2022, Astronomy and Astrophysics, 666, A78

\bibitem[{Daddi {et~al.}(2007)Daddi, Dickinson, Morrison, Chary, Cimatti, Elbaz, Frayer, Renzini, Pope, Alexander, {et~al.}}]{daddi2007MS}
Daddi, E., Dickinson, M., Morrison, G., {et~al.} 2007, The Astrophysical Journal, 670, 156

\bibitem[{De~Barros {et~al.}(2017)De~Barros, Pentericci, Vanzella, Castellano, Fontana, Grazian, Conselice, Yan, Koekemoer, Cristiani, {et~al.}}]{deBarros2017LAEfraction}
De~Barros, S., Pentericci, L., Vanzella, E., {et~al.} 2017, Astronomy \& Astrophysics, 608, A123

\bibitem[{de~La~Vieuville {et~al.}(2019)de~La~Vieuville, Bina, Pello, Mahler, Richard, Drake, Herenz, Bauer, Cl{\'e}ment, Lagattuta, {et~al.}}]{GdlV2019LAELF}
de~La~Vieuville, G., Bina, D., Pello, R., {et~al.} 2019, Astronomy \& Astrophysics, 628, A3

\bibitem[{de~La~Vieuville {et~al.}(2020)de~La~Vieuville, Pell{\'o}, Richard, Mahler, L{\'e}v{\^e}que, Bauer, Lagattuta, Blaizot, Contini, Guaita, {et~al.}}]{GdlV2020LAEfrac}
de~La~Vieuville, G., Pell{\'o}, R., Richard, J., {et~al.} 2020, Astronomy \& Astrophysics, 644, A39

\bibitem[{{Donevski} {et~al.}(2020){Donevski}, {Lapi}, {Ma{\l}ek}, {Liu}, {G{\'o}mez-Guijarro}, {Dav{\'e}}, {Kraljic}, {Pantoni}, {Man}, {Fujimoto}, {Feltre}, {Pearson}, {Li}, \& {Narayanan}}]{Donevski2020dust_stellar_mass}
{Donevski}, D., {Lapi}, A., {Ma{\l}ek}, K., {et~al.} 2020, \aap, 644, A144

\bibitem[{Drake {et~al.}(2017)Drake, Garel, Wisotzki, Leclercq, Hashimoto, Richard, Bacon, Blaizot, Caruana, Conseil, {et~al.}}]{drake2017MUSELAELF}
Drake, A.-B., Garel, T., Wisotzki, L., {et~al.} 2017, Astronomy \& Astrophysics, 608, A6

\bibitem[{Duncan {et~al.}(2014)Duncan, Conselice, Mortlock, Hartley, Guo, Ferguson, Dav{\'e}, Lu, Ownsworth, Ashby, {et~al.}}]{duncan2014masslight}
Duncan, K., Conselice, C.~J., Mortlock, A., {et~al.} 2014, Monthly Notices of the Royal Astronomical Society, 444, 2960

\bibitem[{Elbaz {et~al.}(2007)Elbaz, Daddi, Le~Borgne, Dickinson, Alexander, Chary, Starck, Brandt, Kitzbichler, MacDonald, {et~al.}}]{elbaz2007MS}
Elbaz, D., Daddi, E., Le~Borgne, D., {et~al.} 2007, Astronomy \& Astrophysics, 468, 33

\bibitem[{Elbaz {et~al.}(2011)Elbaz, Dickinson, Hwang, D{\'\i}az-Santos, Magdis, Magnelli, Le~Borgne, Galliano, Pannella, Chanial, {et~al.}}]{elbaz2011goodsMS}
Elbaz, D., Dickinson, M., Hwang, H., {et~al.} 2011, Astronomy \& Astrophysics, 533, A119

\bibitem[{{Ferland} {et~al.}(2017){Ferland}, {Chatzikos}, {Guzm{\'a}n}, {Lykins}, {van Hoof}, {Williams}, {Abel}, {Badnell}, {Keenan}, {Porter}, \& {Stancil}}]{Ferland2017CLOUDY}
{Ferland}, G.~J., {Chatzikos}, M., {Guzm{\'a}n}, F., {et~al.} 2017, \rmxaa, 53, 385

\bibitem[{Finkelstein {et~al.}(2015)Finkelstein, Ryan, Papovich, Dickinson, Song, Somerville, Ferguson, Salmon, Giavalisco, Koekemoer, Ashby, Behroozi, Castellano, Dunlop, Faber, Fazio, Fontana, Grogin, Hathi, Jaacks, Kocevski, Livermore, McLure, Merlin, Mobasher, Newman, Rafelski, Tilvi, \& Willner}]{Finkelstein_2015}
Finkelstein, S.~L., Ryan, R.~E., Papovich, C., {et~al.} 2015, The Astrophysical Journal, 810, 71

\bibitem[{{Fujimoto} {et~al.}(2023){Fujimoto}, {Wang}, {Weaver}, {Kokorev}, {Atek}, {Bezanson}, {Labbe}, {Brammer}, {Greene}, {Chemerynska}, {Dayal}, {de Graaff}, {Furtak}, {Oesch}, {Setton}, {Price}, {Miller}, {Williams}, {Whitaker}, {Zitrin}, {Cutler}, {Leja}, {Pan}, {Coe}, {van Dokkum}, {Feldmann}, {Fudamoto}, {Goulding}, {Khullar}, {Marchesini}, {Maseda}, {Nanayakkara}, {Nelson}, {Smit}, {Stefanon}, \& {Weibel}}]{Fujimoto2023AGNreionisation}
{Fujimoto}, S., {Wang}, B., {Weaver}, J., {et~al.} 2023, arXiv e-prints, arXiv:2308.11609

\bibitem[{Goovaerts {et~al.}(2023)Goovaerts, Pello, Thai, Tuan-Anh, Richard, Claeyssens, Carinos, De~La~Vieuville, \& Matthee}]{goovaerts2023}
Goovaerts, I., Pello, R., Thai, T., {et~al.} 2023, Astronomy \& Astrophysics, 678, A174

\bibitem[{Grazian {et~al.}(2018)Grazian, Giallongo, Boutsia, Cristiani, Vanzella, Scarlata, Santini, Pentericci, Merlin, Menci, {et~al.}}]{Grazian2018AGN}
Grazian, A., Giallongo, E., Boutsia, K., {et~al.} 2018, Astronomy \& Astrophysics, 613, A44

\bibitem[{Hinton(2016)}]{hinton2016marz}
Hinton, S. 2016, Astrophysics Source Code Library, ascl

\bibitem[{{Jiang} {et~al.}(2022){Jiang}, {Ning}, {Fan}, {Ho}, {Luo}, {Wang}, {Wu}, {Wu}, {Yang}, \& {Zheng}}]{JiangAGN2022}
{Jiang}, L., {Ning}, Y., {Fan}, X., {et~al.} 2022, Nature Astronomy, 6, 850

\bibitem[{Jullo {et~al.}(2007)Jullo, Kneib, Limousin, Eliasdottir, Marshall, \& Verdugo}]{jullo2007lenstool}
Jullo, E., Kneib, J.-P., Limousin, M., {et~al.} 2007, New Journal of Physics, 9, 447

\bibitem[{Kneib {et~al.}(1996)Kneib, Ellis, Smail, Couch, \& Sharples}]{kneib1996lenstooloriginal}
Kneib, J.-P., Ellis, R.~S., Smail, I., Couch, W., \& Sharples, R. 1996, The Astrophysical Journal, 471, 643

\bibitem[{{Kokorev} {et~al.}(2022){Kokorev}, {Brammer}, {Fujimoto}, {Kohno}, {Magdis}, {Valentino}, {Toft}, {Oesch}, {Davidzon}, {Bauer}, {Coe}, {Egami}, {Oguri}, {Ouchi}, {Postman}, {Richard}, {Jolly}, {Knudsen}, {Sun}, {Weaver}, {Ao}, {Baker}, {Bradley}, {Caputi}, {Dessauges-Zavadsky}, {Espada}, {Hatsukade}, {Koekemoer}, {Mu{\~n}oz Arancibia}, {Shimasaku}, {Umehata}, {Wang}, \& {Wang}}]{Kokorev2022_A2744HSTdata}
{Kokorev}, V., {Brammer}, G., {Fujimoto}, S., {et~al.} 2022, \apjs, 263, 38

\bibitem[{{Kokorev} {et~al.}(2023){Kokorev}, {Fujimoto}, {Labbe}, {Greene}, {Bezanson}, {Dayal}, {Nelson}, {Atek}, {Brammer}, {Caputi}, {Chemerynska}, {Cutler}, {Feldmann}, {Fudamoto}, {Furtak}, {Goulding}, {de Graaff}, {Leja}, {Marchesini}, {Miller}, {Nanayakkara}, {Oesch}, {Pan}, {Price}, {Setton}, {Smit}, {Stefanon}, {Wang}, {Weaver}, {Whitaker}, {Williams}, \& {Zitrin}}]{Kokorev2023AGN}
{Kokorev}, V., {Fujimoto}, S., {Labbe}, I., {et~al.} 2023, \apjl, 957, L7

\bibitem[{{Kroupa}(2001)}]{Kroupa2001IMF}
{Kroupa}, P. 2001, \mnras, 322, 231

\bibitem[{Kusakabe {et~al.}(2020)Kusakabe, Blaizot, Garel, Verhamme, Bacon, Richard, Hashimoto, Inami, Conseil, Guiderdoni, {et~al.}}]{kusakabe2020}
Kusakabe, H., Blaizot, J., Garel, T., {et~al.} 2020, Astronomy \& Astrophysics, 638, A12

\bibitem[{Laporte {et~al.}(2015)Laporte, Streblyanska, Kim, Pell{\'o}, Bauer, Bina, Brammer, De~Leo, Infante, \& P{\'e}rez-Fournon}]{Laporte2015spitzerlensing}
Laporte, N., Streblyanska, A., Kim, S., {et~al.} 2015, Astronomy \& Astrophysics, 575, A92

\bibitem[{Leclercq {et~al.}(2017)Leclercq, Bacon, Wisotzki, Mitchell, Garel, Verhamme, Blaizot, Hashimoto, Herenz, Conseil, {et~al.}}]{leclercq2017lyahaloes}
Leclercq, F., Bacon, R., Wisotzki, L., {et~al.} 2017, Astronomy \& Astrophysics, 608, A8

\bibitem[{Lee {et~al.}(2010)Lee, Ferguson, Somerville, Wiklind, \& Giavalisco}]{lee2010SEDmass}
Lee, S.-K., Ferguson, H.~C., Somerville, R.~S., Wiklind, T., \& Giavalisco, M. 2010, The Astrophysical Journal, 725, 1644

\bibitem[{Livermore {et~al.}(2017)Livermore, Finkelstein, \& Lotz}]{Livermore_2017}
Livermore, R.~C., Finkelstein, S.~L., \& Lotz, J.~M. 2017, The Astrophysical Journal, 835, 113

\bibitem[{{Looser} {et~al.}(2023){Looser}, {D'Eugenio}, {Maiolino}, {Tacchella}, {Curti}, {Arribas}, {Baker}, {Baum}, {Bonaventura}, {Boyett}, {Bunker}, {Carniani}, {Charlot}, {Chevallard}, {Curtis-Lake}, {Danhaive}, {Eisenstein}, {de Graaff}, {Hainline}, {Ji}, {Johnson}, {Kumari}, {Nelson}, {Parlanti}, {Rix}, {Robertson}, {Rodr{\'\i}guez Del Pino}, {Sandles}, {Scholtz}, {Smit}, {Stark}, {{\"U}bler}, {Williams}, {Willott}, \& {Witstok}}]{Looser2023MS}
{Looser}, T.~J., {D'Eugenio}, F., {Maiolino}, R., {et~al.} 2023, arXiv e-prints, arXiv:2306.02470

\bibitem[{Lotz {et~al.}(2017)Lotz, Koekemoer, Coe, Grogin, Capak, Mack, Anderson, Avila, Barker, Borncamp, {et~al.}}]{lotz2017frontier}
Lotz, J.~e., Koekemoer, A., Coe, D., {et~al.} 2017, The Astrophysical Journal, 837, 97

\bibitem[{Mahler {et~al.}(2018)Mahler, Richard, Cl{\'e}ment, Lagattuta, Schmidt, Patr{\'\i}cio, Soucail, Bacon, Pello, Bouwens, {et~al.}}]{mahler2018strong}
Mahler, G., Richard, J., Cl{\'e}ment, B., {et~al.} 2018, Monthly Notices of the Royal Astronomical Society, 473, 663

\bibitem[{Ma{\l}ek {et~al.}(2018)Ma{\l}ek, Buat, Roehlly, Burgarella, Hurley, Shirley, Duncan, Efstathiou, Papadopoulos, Vaccari, {et~al.}}]{malek2018cigaleSFH}
Ma{\l}ek, K., Buat, V., Roehlly, Y., {et~al.} 2018, Astronomy \& Astrophysics, 620, A50

\bibitem[{{Maseda} {et~al.}(2020){Maseda}, {Bacon}, {Lam}, {Matthee}, {Brinchmann}, {Schaye}, {Labbe}, {Schmidt}, {Boogaard}, {Bouwens}, {Cantalupo}, {Franx}, {Hashimoto}, {Inami}, {Kusakabe}, {Mahler}, {Nanayakkara}, {Richard}, \& {Wisotzki}}]{Maseda2020IRAC_Halpha_contam}
{Maseda}, M.~V., {Bacon}, R., {Lam}, D., {et~al.} 2020, \mnras, 493, 5120

\bibitem[{{Maseda} {et~al.}(2023){Maseda}, {Lewis}, {Matthee}, {Hennawi}, {Boogaard}, {Feltre}, {Nanayakkara}, {Bacon}, {Barger}, {Brinchmann}, {Franx}, {Hashimoto}, {Inami}, {Kusakabe}, {Leclercq}, {Rowland}, {Taylor}, {Tremonti}, {Urrutia}, {Schaye}, {Simmonds}, \& {Vitte}}]{Maseda2023NIRSpec_Halpha}
{Maseda}, M.~V., {Lewis}, Z., {Matthee}, J., {et~al.} 2023, arXiv e-prints, arXiv:2304.08511

\bibitem[{{Matthee} {et~al.}(2023){Matthee}, {Mackenzie}, {Simcoe}, {Kashino}, {Lilly}, {Bordoloi}, \& {Eilers}}]{Matthee2023JWST_nebular_lines}
{Matthee}, J., {Mackenzie}, R., {Simcoe}, R.~A., {et~al.} 2023, \apj, 950, 67

\bibitem[{Matthee {et~al.}(2022)Matthee, Naidu, Pezzulli, Gronke, Sobral, Oesch, Hayes, Erb, Schaerer, Amor{\'\i}n, {et~al.}}]{Mathee2022brightlya}
Matthee, J., Naidu, R.~P., Pezzulli, G., {et~al.} 2022, Monthly Notices of the Royal Astronomical Society, 512, 5960

\bibitem[{{Matthee} {et~al.}(2021){Matthee}, {Sobral}, {Hayes}, {Pezzulli}, {Gronke}, {Schaerer}, {Naidu}, {R{\"o}ttgering}, {Calhau}, {Paulino-Afonso}, {Santos}, \& {Amor{\'\i}n}}]{Matthee2022what_makes_LAE}
{Matthee}, J., {Sobral}, D., {Hayes}, M., {et~al.} 2021, \mnras, 505, 1382

\bibitem[{McGreer {et~al.}(2015)McGreer, Mesinger, \& D'Odorico}]{mcgreer2015reionisation}
McGreer, I.~D., Mesinger, A., \& D'Odorico, V. 2015, Monthly Notices of the Royal Astronomical Society, 447, 499

\bibitem[{Mercier {et~al.}(2022)Mercier, Epinat, Contini, Abril-Melgarejo, Boogaard, Brinchmann, Finley, Krajnovi{\'c}, Michel-Dansac, Ventou, {et~al.}}]{mercier2022MAGIC_SEDfitting}
Mercier, W., Epinat, B., Contini, T., {et~al.} 2022, Astronomy \& Astrophysics, 665, A54

\bibitem[{Nakane {et~al.}(2023)Nakane, Ouchi, Nakajima, Harikane, Ono, Umeda, Isobe, Zhang, \& Xu}]{Nakane2023highzLAEs}
Nakane, M., Ouchi, M., Nakajima, K., {et~al.} 2023 [\eprint{Arxiv:2312.06804v1}]

\bibitem[{Pentericci {et~al.}(2011)Pentericci, Fontana, Vanzella, Castellano, Grazian, Dijkstra, Boutsia, Cristiani, Dickinson, Giallongo, {et~al.}}]{pentericci2011laefrac/z=7LBG}
Pentericci, L., Fontana, A., Vanzella, E., {et~al.} 2011, The Astrophysical Journal, 743, 132

\bibitem[{Pentericci {et~al.}(2018)Pentericci, Vanzella, Castellano, Fontana, De~Barros, Grazian, Marchi, Bradac, Conselice, Cristiani, {et~al.}}]{pentericci2018LAEfrac}
Pentericci, L., Vanzella, E., Castellano, M., {et~al.} 2018, Astronomy \& Astrophysics, 619, A147

\bibitem[{Piqueras {et~al.}(2017)Piqueras, Conseil, Shepherd, Bacon, Leclercq, \& Richard}]{piqueras2017mpdaf}
Piqueras, L., Conseil, S., Shepherd, M., {et~al.} 2017, arXiv preprint arXiv:1710.03554

\bibitem[{{Planck Collaboration} {et~al.}(2020){Planck Collaboration}, {Aghanim, N.}, {Akrami, Y.}, {Ashdown, M.}, {Aumont, J.}, {Baccigalupi, C.}, {Ballardini, M.}, {Banday, A. J.}, {Barreiro, R. B.}, {Bartolo, N.}, {Basak, S.}, {Battye, R.}, {Benabed, K.}, {Bernard, J.-P.}, {Bersanelli, M.}, {Bielewicz, P.}, {Bock, J. J.}, {Bond, J. R.}, {Borrill, J.}, {Bouchet, F. R.}, {Boulanger, F.}, {Bucher, M.}, {Burigana, C.}, {Butler, R. C.}, {Calabrese, E.}, {Cardoso, J.-F.}, {Carron, J.}, {Challinor, A.}, {Chiang, H. C.}, {Chluba, J.}, {Colombo, L. P. L.}, {Combet, C.}, {Contreras, D.}, {Crill, B. P.}, {Cuttaia, F.}, {de Bernardis, P.}, {de Zotti, G.}, {Delabrouille, J.}, {Delouis, J.-M.}, {Di Valentino, E.}, {Diego, J. M.}, {Dor\'e, O.}, {Douspis, M.}, {Ducout, A.}, {Dupac, X.}, {Dusini, S.}, {Efstathiou, G.}, {Elsner, F.}, {En\ss{}lin, T. A.}, {Eriksen, H. K.}, {Fantaye, Y.}, {Farhang, M.}, {Fergusson, J.}, {Fernandez-Cobos, R.}, {Finelli, F.}, {Forastieri, F.}, {Frailis, M.}, {Fraisse, A. A.}, {Franceschi, E.},
  {Frolov, A.}, {Galeotta, S.}, {Galli, S.}, {Ganga, K.}, {G\'enova-Santos, R. T.}, {Gerbino, M.}, {Ghosh, T.}, {Gonz\'alez-Nuevo, J.}, {G\'orski, K. M.}, {Gratton, S.}, {Gruppuso, A.}, {Gudmundsson, J. E.}, {Hamann, J.}, {Handley, W.}, {Hansen, F. K.}, {Herranz, D.}, {Hildebrandt, S. R.}, {Hivon, E.}, {Huang, Z.}, {Jaffe, A. H.}, {Jones, W. C.}, {Karakci, A.}, {Keih\"anen, E.}, {Keskitalo, R.}, {Kiiveri, K.}, {Kim, J.}, {Kisner, T. S.}, {Knox, L.}, {Krachmalnicoff, N.}, {Kunz, M.}, {Kurki-Suonio, H.}, {Lagache, G.}, {Lamarre, J.-M.}, {Lasenby, A.}, {Lattanzi, M.}, {Lawrence, C. R.}, {Le Jeune, M.}, {Lemos, P.}, {Lesgourgues, J.}, {Levrier, F.}, {Lewis, A.}, {Liguori, M.}, {Lilje, P. B.}, {Lilley, M.}, {Lindholm, V.}, {L\'opez-Caniego, M.}, {Lubin, P. M.}, {Ma, Y.-Z.}, {Mac\'{\i}as-P\'erez, J. F.}, {Maggio, G.}, {Maino, D.}, {Mandolesi, N.}, {Mangilli, A.}, {Marcos-Caballero, A.}, {Maris, M.}, {Martin, P. G.}, {Martinelli, M.}, {Mart\'{\i}nez-Gonz\'alez, E.}, {Matarrese, S.}, {Mauri, N.}, {McEwen, J. D.},
  {Meinhold, P. R.}, {Melchiorri, A.}, {Mennella, A.}, {Migliaccio, M.}, {Millea, M.}, {Mitra, S.}, {Miville-Desch\^enes, M.-A.}, {Molinari, D.}, {Montier, L.}, {Morgante, G.}, {Moss, A.}, {Natoli, P.}, {N\o{}rgaard-Nielsen, H. U.}, {Pagano, L.}, {Paoletti, D.}, {Partridge, B.}, {Patanchon, G.}, {Peiris, H. V.}, {Perrotta, F.}, {Pettorino, V.}, {Piacentini, F.}, {Polastri, L.}, {Polenta, G.}, {Puget, J.-L.}, {Rachen, J. P.}, {Reinecke, M.}, {Remazeilles, M.}, {Renzi, A.}, {Rocha, G.}, {Rosset, C.}, {Roudier, G.}, {Rubi\~no-Mart\'{\i}n, J. A.}, {Ruiz-Granados, B.}, {Salvati, L.}, {Sandri, M.}, {Savelainen, M.}, {Scott, D.}, {Shellard, E. P. S.}, {Sirignano, C.}, {Sirri, G.}, {Spencer, L. D.}, {Sunyaev, R.}, {Suur-Uski, A.-S.}, {Tauber, J. A.}, {Tavagnacco, D.}, {Tenti, M.}, {Toffolatti, L.}, {Tomasi, M.}, {Trombetti, T.}, {Valenziano, L.}, {Valiviita, J.}, {Van Tent, B.}, {Vibert, L.}, {Vielva, P.}, {Villa, F.}, {Vittorio, N.}, {Wandelt, B. D.}, {Wehus, I. K.}, {White, M.}, {White, S. D. M.}, {Zacchei, A.}, \&
  {Zonca, A.}}]{Planck2018reionisation}
{Planck Collaboration}, {Aghanim, N.}, {Akrami, Y.}, {et~al.} 2020, A\&A, 641, A6

\bibitem[{Popesso {et~al.}(2023)Popesso, Concas, Cresci, Belli, Rodighiero, Inami, Dickinson, Ilbert, Pannella, \& Elbaz}]{P23MS}
Popesso, P., Concas, A., Cresci, G., {et~al.} 2023, Monthly Notices of the Royal Astronomical Society, 519, 1526

\bibitem[{Pozzetti {et~al.}(2010)Pozzetti, Bolzonella, Zucca, Zamorani, Lilly, Renzini, Moresco, Mignoli, Cassata, Tasca, {et~al.}}]{pozzetti2010SEDfittingmass}
Pozzetti, L., Bolzonella, M., Zucca, E., {et~al.} 2010, Astronomy \& Astrophysics, 523, A13

\bibitem[{Richard {et~al.}(2021)Richard, Claeyssens, Lagattuta, Guaita, Bauer, Pello, Carton, Bacon, Soucail, Lyon, {et~al.}}]{richard2021atlas}
Richard, J., Claeyssens, A., Lagattuta, D., {et~al.} 2021, Astronomy \& Astrophysics, 646, A83

\bibitem[{Salpeter(1955)}]{salpeter1955IMF}
Salpeter, E.~E. 1955, The Astrophysical Journal, 121, 161

\bibitem[{Santini {et~al.}(2017)Santini, Fontana, Castellano, Di~Criscienzo, Merlin, Amorin, Cullen, Daddi, Dickinson, Dunlop, {et~al.}}]{S17MS}
Santini, P., Fontana, A., Castellano, M., {et~al.} 2017, The Astrophysical Journal, 847, 76

\bibitem[{Santini {et~al.}(2023)Santini, Fontana, Castellano, Leethochawalit, Trenti, Treu, Belfiori, Birrer, Bonchi, Merlin, {et~al.}}]{santini2023masslightGLASS}
Santini, P., Fontana, A., Castellano, M., {et~al.} 2023, The Astrophysical Journal Letters, 942, L27

\bibitem[{Santini {et~al.}(2009)Santini, Fontana, Grazian, Salimbeni, Fiore, Fontanot, Boutsia, Castellano, Cristiani, De~Santis, {et~al.}}]{santini2009MS}
Santini, P., Fontana, A., Grazian, A., {et~al.} 2009, Astronomy \& Astrophysics, 504, 751

\bibitem[{Santos {et~al.}(2021)Santos, Sobral, Butterworth, Paulino-Afonso, Ribeiro, Da~Cunha, Calhau, Khostovan, Matthee, \& Arrabal~Haro}]{santos2021LAEmassrelation}
Santos, S., Sobral, D., Butterworth, J., {et~al.} 2021, Monthly Notices of the Royal Astronomical Society, 505, 1117

\bibitem[{Santos {et~al.}(2020)Santos, Sobral, Matthee, Calhau, Da~Cunha, Ribeiro, Paulino-Afonso, Arrabal~Haro, \& Butterworth}]{santos2020SC4K_LAE_MS}
Santos, S., Sobral, D., Matthee, J., {et~al.} 2020, Monthly Notices of the Royal Astronomical Society, 493, 141

\bibitem[{{Schaerer} {et~al.}(2022){Schaerer}, {Marques-Chaves}, {Barrufet}, {Oesch}, {Izotov}, {Naidu}, {Guseva}, \& {Brammer}}]{Schaerer2022JWST_high_z_galaxies}
{Schaerer}, D., {Marques-Chaves}, R., {Barrufet}, L., {et~al.} 2022, \aap, 665, L4

\bibitem[{Shipley {et~al.}(2018)Shipley, Lange-Vagle, Marchesini, Brammer, Ferrarese, Stefanon, Kado-Fong, Whitaker, Oesch, Feinstein, {et~al.}}]{shipley2018hff}
Shipley, H.~V., Lange-Vagle, D., Marchesini, D., {et~al.} 2018, The Astrophysical Journal Supplement Series, 235, 14

\bibitem[{Sobral {et~al.}(2014)Sobral, Best, Smail, Mobasher, Stott, \& Nisbet}]{sobral2014stellar}
Sobral, D., Best, P.~N., Smail, I., {et~al.} 2014, Monthly Notices of the Royal Astronomical Society, 437, 3516

\bibitem[{Speagle {et~al.}(2014)Speagle, Steinhardt, Capak, \& Silverman}]{S14MS}
Speagle, J.~S., Steinhardt, C.~L., Capak, P.~L., \& Silverman, J.~D. 2014, The Astrophysical Journal Supplement Series, 214, 15

\bibitem[{Stark {et~al.}(2010)Stark, Ellis, Chiu, Ouchi, \& Bunker}]{stark2010keckLAEfrac}
Stark, D.~P., Ellis, R.~S., Chiu, K., Ouchi, M., \& Bunker, A. 2010, Monthly Notices of the Royal Astronomical Society, 408, 1628

\bibitem[{Stark {et~al.}(2011)Stark, Ellis, \& Ouchi}]{stark2011LAEfrac}
Stark, D.~P., Ellis, R.~S., \& Ouchi, M. 2011, The Astrophysical Journal Letters, 728, L2

\bibitem[{Steinhardt {et~al.}(2014)Steinhardt, Speagle, Capak, Silverman, Carollo, Dunlop, Hashimoto, Hsieh, Ilbert, Le~Fevre, {et~al.}}]{steinhardt2014MS}
Steinhardt, C.~L., Speagle, J.~S., Capak, P., {et~al.} 2014, The Astrophysical journal letters, 791, L25

\bibitem[{Thai {et~al.}(2023)Thai, Tuan-Anh, Pello, Goovaerts, Richard, Claeyssens, Mahler, Lagattuta, de~la Vieuville, Salvador-Sol{\'e}, {et~al.}}]{Thai2023LF}
Thai, T.~T., Tuan-Anh, P., Pello, R., {et~al.} 2023, Astronomy \& Astrophysics, 678, A139

\bibitem[{Verhamme {et~al.}(2008)Verhamme, Schaerer, Atek, \& Tapken}]{verhamme2008lyadustsim}
Verhamme, A., Schaerer, D., Atek, H., \& Tapken, C. 2008, Astronomy \& Astrophysics, 491, 89

\bibitem[{Weaver {et~al.}(2023)Weaver, Cutler, Pan, Whitaker, Labbe, Price, Bezanson, Brammer, Marchesini, Leja, {et~al.}}]{weaver2023UNCOVERcatalogs}
Weaver, J.~R., Cutler, S.~E., Pan, R., {et~al.} 2023, arXiv preprint arXiv:2301.02671

\bibitem[{{Zahid} {et~al.}(2012){Zahid}, {Dima}, {Kewley}, {Erb}, \& {Dav{\'e}}}]{Zahid2012IMFcorrections}
{Zahid}, H.~J., {Dima}, G.~I., {Kewley}, L.~J., {Erb}, D.~K., \& {Dav{\'e}}, R. 2012, \apj, 757, 54

\bibitem[{Zahid {et~al.}(2013)Zahid, Yates, Kewley, \& Kudritzki}]{zahid2013stellar_mass_dust}
Zahid, H.~J., Yates, R.~M., Kewley, L.~J., \& Kudritzki, R.-P. 2013, The Astrophysical Journal, 763, 92

\end{thebibliography}

\begin{appendix}
\section{\cg Fitting Details}
\label{sect:app_cg}
\begin{figure}[h]
    \centering
    \includegraphics[width=0.5\textwidth]{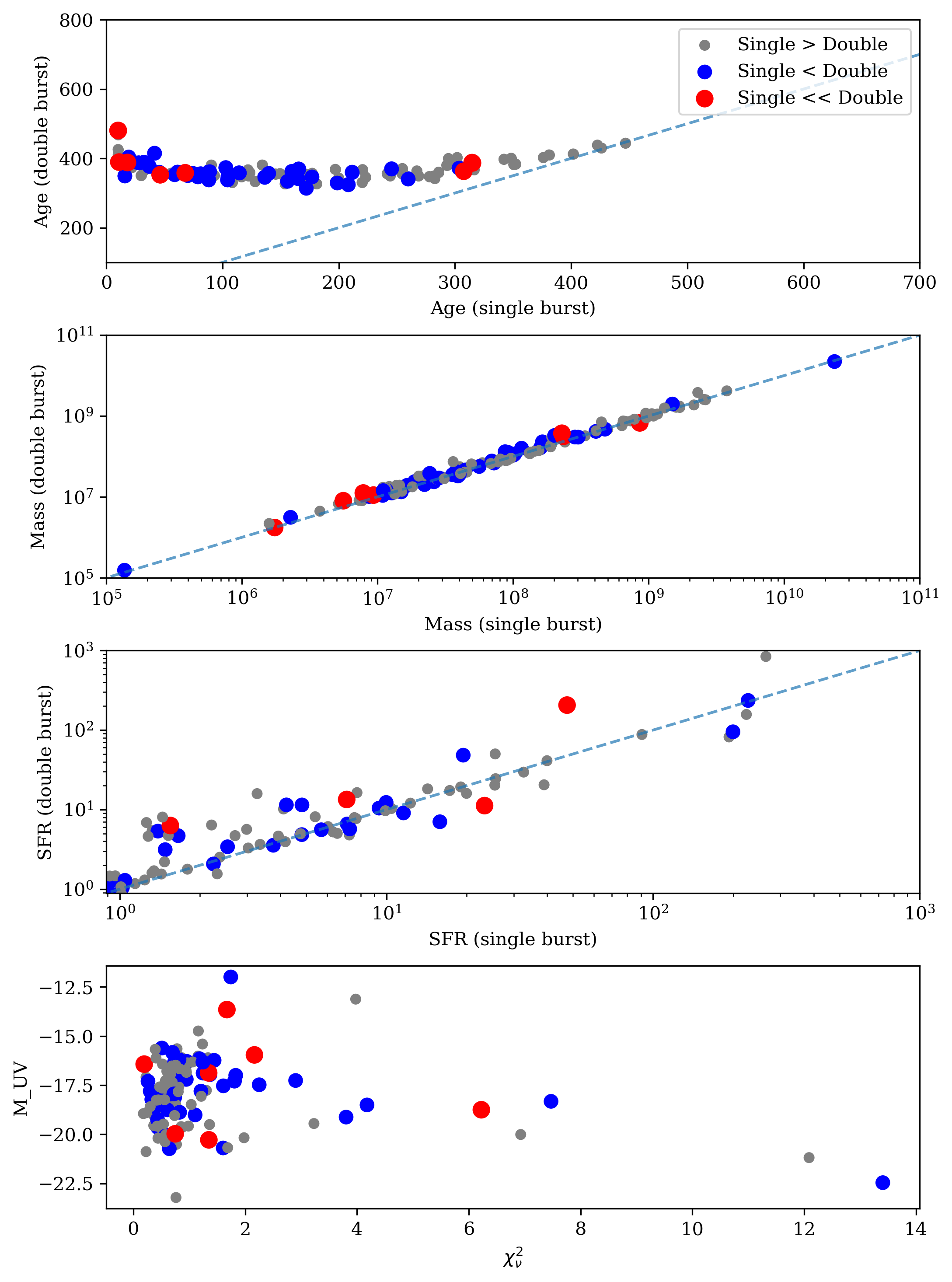}
    \caption{Comparison of the two SFHs used in the \cg fitting process: the single burst model (x-axis) and the double burst model (y-axis). These graphs include the highly-magnified object described in the main text and removed from the MS fitting. In descending order, the graphs depict the comparison of the age of the main stellar population, the stellar mass, the SFR and the $\chi^2_{\nu}$ statistic. Grey dots indicate galaxies for which the single burst model is a better fit and coloured dots indicate galaxies for which the double burst model is better. Red dots indicate galaxies for which the double burst model is significantly better than the single burst model. Error bars on these values are described in the caption of Fig.~\ref{fig:MS}.}
    
    \label{fig:cg_SFH}
\end{figure}
\begin{figure}
    \centering
    \includegraphics[width=0.5\textwidth]{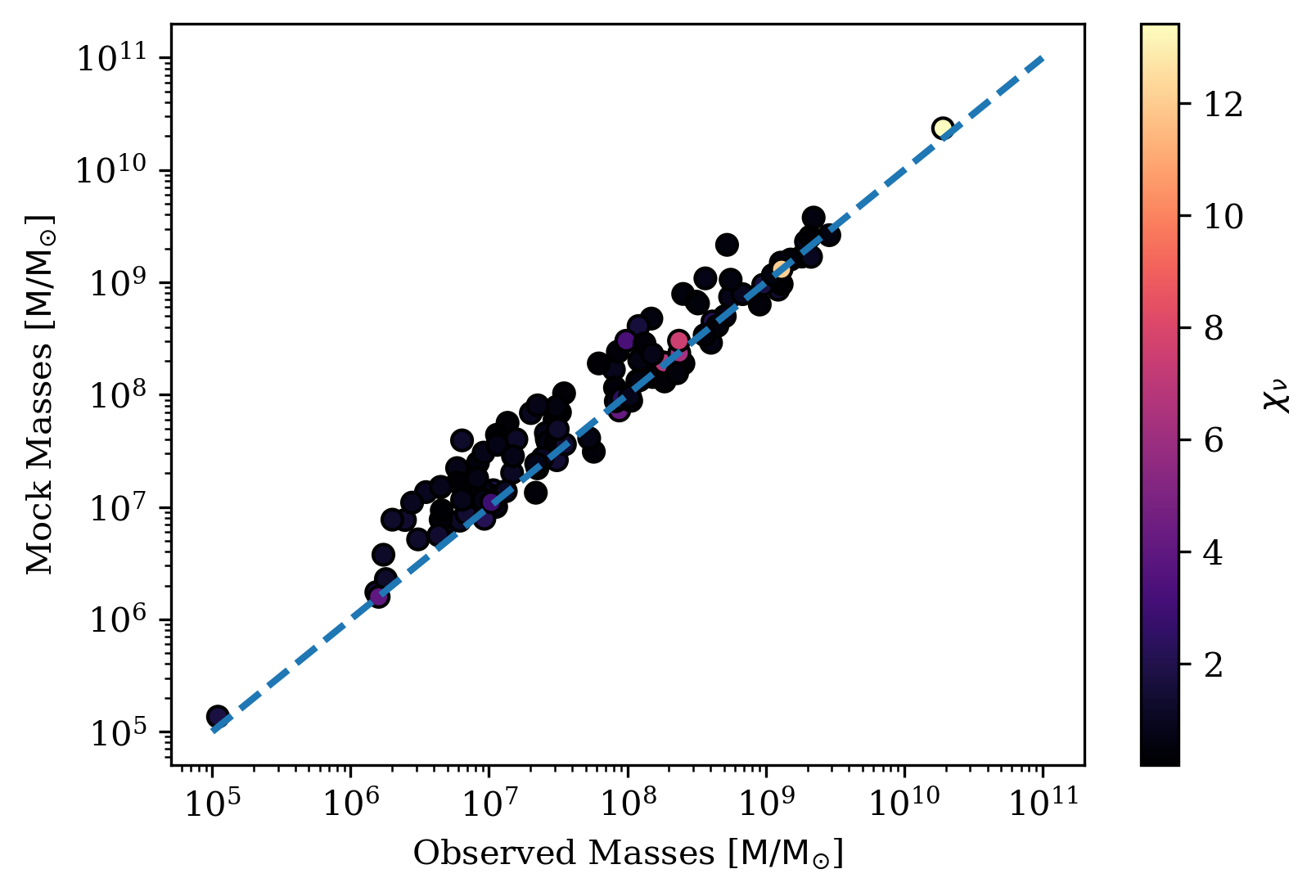}
    \\[\smallskipamount]
    \includegraphics[width=0.5\textwidth]{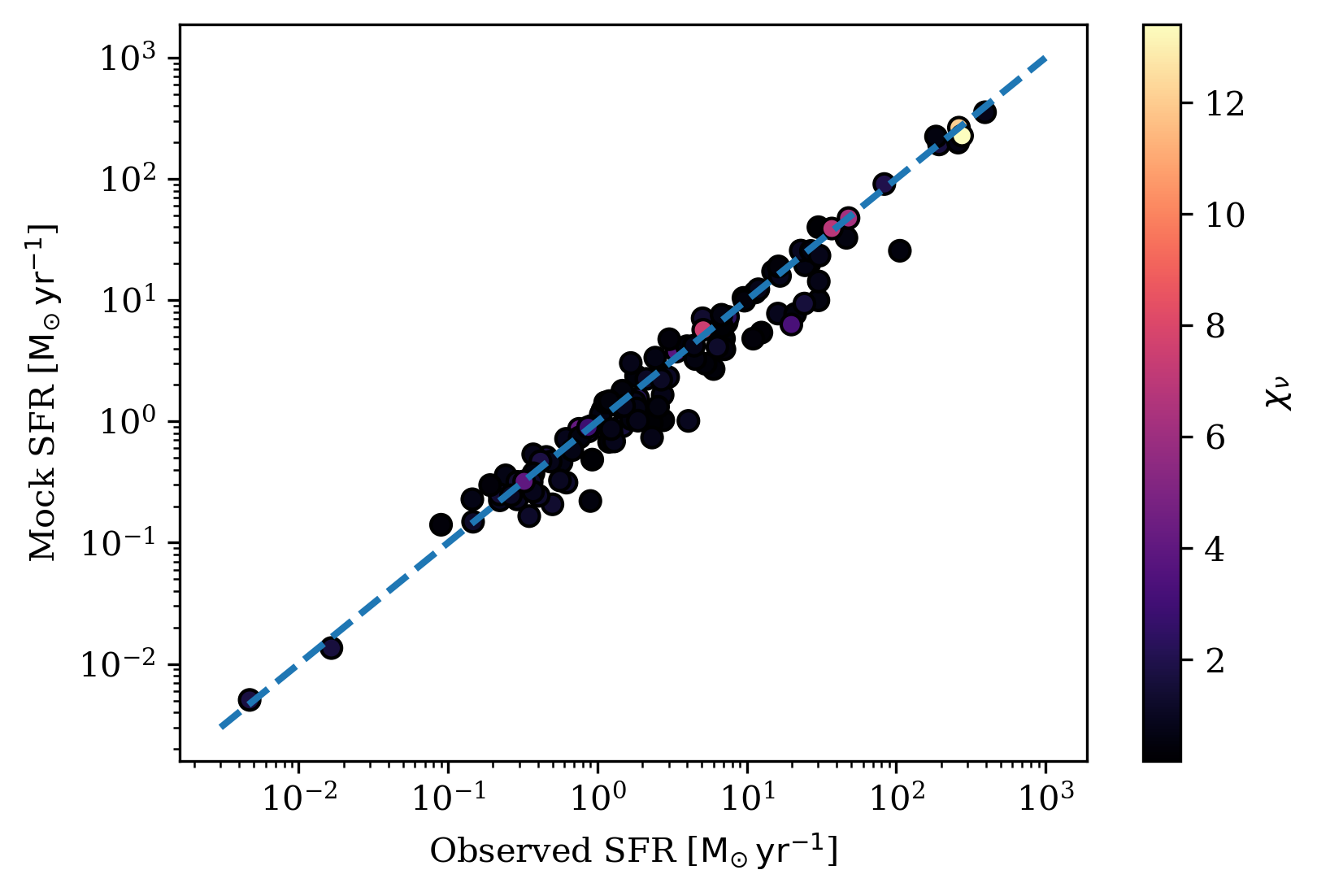}
    \caption{The results of the mock catalog creation and fitting.\\
    \textit{Upper:} Comparison of observed and mock stellar masses, colour-coded by the $\chi^2_{\nu}$ statistic of the original, observed fit. \\
    \textit{Lower:} Comparison of observed and mock SFRs, colour-coded by the $\chi^2_{\nu}$ statistic of the original, observed fit.
    }
    
    \label{fig:cgmock}
\end{figure}

\indent Here we give further details of the SED fitting we performed with \cg, in particular, the checks we performed to ensure that this fitting process results in reliable stellar mass and SFR estimates. \\
\indent First, we depict in Fig.~\ref{fig:cg_SFH}, the comparison (in terms of resulting age, stellar mass, SFR and $\chi^2_{\nu}$), between the two SFHs used in the \cg fitting. The best-fit SFH is chosen moving forward from the SED fitting stage. The graphs depicting stellar mass and SFR demonstrate how robust the SED fitting process is to changes in SFH. Particularly the stellar mass changes very little between SFHs. This has already been noted by \cite{pozzetti2010SEDfittingmass,Bolzonella2010SEDfittingmass,lee2010SEDmass,ciesla2017sfrmass} among others, however, see also \cite{buat2014SED}. We note that the addition of \textit{JWST}/NIRCam data into the near-infrared helps mass determination at these redshifts significantly compared to previously available data.\\
\indent The SFR exhibits a higher dispersion when comparing the two SFHs, however, there is no significant offset. This greater dispersion in the SED-derived SFR is expected, see the above references, as well as \cite{mercier2022MAGIC_SEDfitting}. This comes from the lack of data in the rest frame mid and far infrared, which would be useful to constrain the dust content of these galaxies (although it is expected to be limited \citep{deBarros2017LAEfraction,goovaerts2023}).\\
\indent A further check on the reliability of galaxy parameter estimation is performed using \cg's mock catalog function. This is a self-consistency check on derived physical parameters, inbuilt within \cg, which uses the best-fit parameters as truth values, creating an artificial catalog to which noise, drawn from the observed uncertainties, is added. These artificial observations then undergo the same fitting process as the actual observations, and a comparison of the artificial and best-fit parameters gives an indication of how well the physical quantities are retrieved. This comparison is shown in Fig.~\ref{fig:cgmock}.\\
\indent Altogether, these tests show that \cg can reliably recover the relevant physical parameters of the LAEs for the estimation of the MS relation. \\

\end{appendix}

\end{document}